\PassOptionsToPackage{svgnames}{xcolor}
\documentclass[12pt,a4paper]{article}
\usepackage{tikz}
\usepackage{pgf}

\usepackage[english]{babel}

\usepackage{amsmath}
\usepackage{booktabs}
\usepackage{cite}
\usepackage[colorlinks=true,linkcolor=black, citecolor=black,urlcolor=black]{hyperref}

\usepackage{multirow,graphics}
\usepackage{enumerate}
\usepackage{amstext}
\usepackage{amssymb}
\usepackage{amsmath}
\usepackage{slashed}

\usepackage{url}
\usepackage{graphicx}
\usepackage{subcaption}
\usepackage{color}
\usepackage[nodisplayskipstretch]{setspace}
\usepackage{tcolorbox}
\usepackage{float}
\usepackage{wasysym}
\usepackage{ytableau}
\usepackage{textcomp}
\usepackage[perpage]{footmisc}

\usetikzlibrary{decorations.pathmorphing}
\usetikzlibrary{decorations.markings}
\usetikzlibrary{intersections}
\usetikzlibrary{calc}
\usetikzlibrary{positioning}
\usetikzlibrary{3d}
\usetikzlibrary{shapes}
\usetikzlibrary{angles}
\usetikzlibrary{fit}
\usetikzlibrary{backgrounds}
\usetikzlibrary{shadows}
\tikzset{
photon/.style={decorate, decoration={snake}},
particle/.style={postaction={decorate},
    decoration={markings,mark=at position .5 with {\arrow{>}}}},
antiparticle/.style={postaction={decorate},
    decoration={markings,mark=at position .5 with {\arrow{<}}}},
gluon/.style={decorate, decoration={coil,amplitude=2pt, segment length=4pt},color=purple},
wilson/.style={color=blue, thick},
scalarZ/.style={postaction={decorate},decoration={markings, mark=at position .5 with{\arrow[scale=1]{stealth}}}},
scalarX/.style={postaction={decorate}, dashed, dash pattern = on 4pt off 2pt, dash phase = 2pt, decoration={markings, mark=at position .53 with{\arrow[scale=1]{stealth}}}},
scalarZw/.style={postaction={decorate},decoration={markings, mark=at position .75 with{\arrow[scale=1]{stealth}}}},
scalarXw/.style={postaction={decorate}, dashed, dash pattern = on 4pt off 2pt, dash phase = 2pt, decoration={markings, mark=at position .60 with{\arrow[scale=1]{stealth}}}},
frozen/.style={inner sep=0.7mm, rectangle,draw},
frozenblue/.style={rectangle, draw, fill=blue!20, inner sep=0.7mm},
norm/.style={->, draw, shorten <=2pt, shorten >=2pt},
diag/.style={->, draw, shorten <=5pt, shorten >=3pt},
every node/.style={inner sep=0.5mm},
webarrow/.style={postaction={decorate},,decoration={markings, mark=at position .5 with{\arrow[scale=1]{stealth}}}}
}

 \newtheorem{Corollary}{Corollary}[section]
 
\makeatletter
\tikzoption{canvas is plane}[]{\@setOxy#1}
\def\@setOxy O(#1,#2,#3)x(#4,#5,#6)y(#7,#8,#9)%
  {\def\tikz@plane@origin{\pgfpointxyz{#1}{#2}{#3}}%
   \def\tikz@plane@x{\pgfpointxyz{#4}{#5}{#6}}%
   \def\tikz@plane@y{\pgfpointxyz{#7}{#8}{#9}}%
   \tikz@canvas@is@plane
  }
\makeatother

\newtheorem{Theorem}{Theorem}[section]
\newtheorem{Example}{Example}[section]

\newcommand\be{\begin{equation}}
\newcommand\ee{\end{equation}}

\normalsize
\makeatletter
\divide\@tempdima\baselineskip
\@tempcnta=\@tempdima
\skip\@mpfootins = \skip\footins
\renewcommand{\@dotsep}{10000}
\makeatother

\begin{document}
\numberwithin{equation}{section}
\begin{center}
\phantom{vv}

\vspace{3cm}
\bigskip

{\Large \bf Adjacency for scattering amplitudes from the Gr\"obner fan}

\bigskip
\mbox{\bf L. Bossinger${}^1$, J. M. Drummond${}^2$, R. Glew${}^2$}%

\bigskip

  ${}^1${\em Instituto de Matem\'aticas, Unidad Oaxaca, Universidad Nacional Aut\'onoma de M\'exico, Le\'on 2, 68000 Oaxaca, Mexico}

${}^2${\em School of Physics \& Astronomy, University of Southampton,\\
  Highfield, Southampton, SO17 1BJ, United Kingdom.}\\[10pt]

\vspace{1cm}  {\bf Abstract}
\end{center}

Scattering amplitudes in planar $\mathcal{N}=4$ super Yang-Mills theory exhibit singularities which reflect various aspects of the cluster algebras associated to the Grassmannians ${\rm Gr}(4,n)$ and their tropical counterparts. Here we investigate the potential origins of such structures and examine the extent to which they can be recovered from the Gr\"obner structure of the underlying Pl\"ucker ideals, focussing on the Grassmannians corresponding to finite cluster algebras.

Starting from the Pl\"ucker ideal, we describe how the polynomial cluster variables are encoded in non-prime initial ideals associated to certain maximal cones of the positive tropical fan. Following \cite{boss2021grob} we show that extending the Pl\"ucker ideal by such variables leads to a Gr\"obner fan with a single maximal Gr\"obner cone spanned by the positive tropical rays. The associated initial ideal encodes the compatibility relations among the full set of cluster variables. Thus we find that the Gr\"obner structure naturally encodes both the symbol alphabet and the cluster adjacency relations exhibited by scattering amplitudes without invoking the cluster algebra at all.

As a potential application of these ideas we then examine the kinematic ideal associated to non-dual conformal massless scattering written in terms of spinor helicity variables. For five-particle scattering we find that the ideal can be identified with the Pl\"ucker ideal for ${\rm Gr}(3,6)$ and the corresponding tropical fan contains a number of non-prime ideals which encode all additional letters of the two-loop pentagon function alphabet present in various calculations of massless five-point finite remainders.

\noindent

\newpage
\phantom{vv}
\vspace{1cm}
\hrule
\tableofcontents

\bigskip
\medskip

\hrule

\section{Introduction}\setcounter{equation}{0}
The connection between the branch cut structure of perturbative $n$-point scattering amplitudes in planar $\mathcal{N}=4$ super Yang-Mills theory and the cluster algebras associated to the Grassmannians ${\rm Gr}(4,n)$ is by now well established \cite{Golden:2013xva,Golden:2014xqa,Goncharov:2010jf}\footnote{The notion of cluster algebras was originally developed in \cite{1021.16017,1054.17024,fomin_zelevinsky_2007}}. Such amplitudes can be expressed (at least for small enough $n$ and loop order) as generalised polylogarithmic functions, a class of iterated integrals built upon a specified symbol alphabet $\mathbb{A}$. The set of cluster variables of the Grassmannian cluster algebras ${\rm Gr}(4,6)$ and ${\rm Gr}(4,7)$ explain the $9$ and $42$ letter alphabets for the hexagon and heptagon amplitudes respectively, the knowledge of which has since facilitated calculations in the context of the analytic bootstrap up to high loop orders \cite{Dixon:2011pw,Dixon:2013eka,Dixon:2014voa,Dixon:2014iba,Dixon:2015iva,Caron-Huot:2016owq,Caron-Huot:2019vjl,Drummond:2014ffa,Dixon:2016nkn,Drummond:2018caf}. Furthermore, the link between the cluster algebras and the singularities of the amplitudes was deepened with the discovery of {\it cluster adjacency} \cite{Drummond:2017ssj,Drummond:2018dfd} which states that consecutive singularities can only appear in the symbol if there exists a cluster in which both letters are contained. This notion of cluster adjacency is closely related to the Steinmann relations, utilised by the bootstrap programme, and their extended counterparts \cite{Caron-Huot:2016owq,Caron-Huot:2019bsq}. 

Recently, efforts have been made to explore the connection between amplitudes and cluster algebras at eight points and beyond where the respective Grassmannian cluster algebra ${\rm Gr}(4,8)$ is no longer of finite type. Two complications arise in this instance. Firstly, some truncation of the infinite set of cluster coordinates must be introduced in order to obtain a finite symbol alphabet and secondly, we must find a way of extracting algebraic square root letters known to appear in the symbol alphabet at eight points, for example from the four-mass one-loop box integral present in N${}^2$MHV amplitudes. 
This has led to the study of the eight-point case and related problems from many closely related approaches including tropical geometry, plabic graphs, Schubert problems and more \cite{Drummond:2019qjk,Drummond:2019cxm,Arkani-Hamed:2019rds,Henke:2019hve,Herderschee:2021dez,Henke:2021ity,Yang:2022gko,He:2022tph,He:2022ujv}. Notably \cite{Drummond:2019cxm}, by assigning expressions to certain rays appearing in the tropicalisation of ${\rm Gr}(4,8)$, which can also be obtained via infinite sequences of mutations in the cluster algebra, one obtains the set of $18$ multiplicatively independent square root letters appearing in the two-loop NMHV amplitude found in \cite{He:2019jee}.

Since the tropicalisation of a polynomial ideal is a subset of the Gr\"obner fan of the ideal by the fundamental theorem of tropical geometry \cite[\S3.2]{SturmfelsMaclaganBook}, the connection between the cluster algebra and the tropical Grassmannian leads us to ask whether the salient features of the cluster algebra can be obtained by considering the Gr\"obner structure of the underlying ideal generated by the Pl\"ucker relations. 
While connections between tropical Grassmannians \cite{SpeyerSturmfels} and cluster algebras had been discussed already in e.g. \cite{speyer2003tropical,brodsky2015cluster}, it is only recently that the relationship between cluster algebras and Gr\"obner theory has been investigated in detail \cite{boss2021grob,Ilten-Najera-Treff}. We shall see, using methods developed in these references, that both the alphabet and the adjacency relations can indeed be obtained from the Gr\"obner structure of the Pl\"ucker ideal or an appropriate extension thereof.

In the case of ${\rm Gr}(2,n)$, all cluster variables are Pl\"ucker variables. For the finite cases ${\rm Gr}(3,n)$ for $n=6,7,8$, we will see that cluster variables which are polynomial in Pl\"ucker coordinates are associated to certain non-prime initial ideals associated to maximal cones of the positive tropical fan $\text{Trop}^+(I_{k,n})$ (a subfan of the Gr\"obner fan). Furthermore, by extending the Pl\"ucker ideal by this additional set of cluster variables, a particular maximal cone of the Gr\"obner fan will be singled out by the rays of $\text{Trop}^+(I^{\text{ext}}_{k,n})$.  The associated monomial initial ideal provides the set of {\it forbidden pairs} of cluster variables i.e. variables which do not appear together in a cluster. For the cases of ${\rm Gr}(2,6) \cong {\rm Gr}(4,6)$ and ${\rm Gr}(3,7) \cong {\rm Gr}(4,7)$ this provides the information required for the amplitude bootstrap programme in the form of the symbol alphabet $\mathbb{A}$ and adjacency rules for the symbol of the amplitude.

Motivated by the fact that interesting physical information can be obtained from the Gr\"obner structure of the kinematic ideal for amplitudes in planar $\mathcal{N}=4$ super Yang-Mills theory, we then consider the kinematics of non-planar and non-dual conformal massless scattering described in spinor helicity variables. We consider an ideal generated by the polynomial relations satisfied by the spinor brackets $\langle ij \rangle$ and $[ij]$ and investigate its Gr\"obner structure. In particular we consider the ideal $I_{5\text{pt}}$ corresponding to the case of scattering five massless particles. In fact we observe that, upon an appropriate identification of variables, this ideal can be identified with the Pl\"ucker ideal for ${\rm Gr}(3,6)$. By searching for non-prime initial ideals, this time in the full tropical space ${\rm Trop}(I_{5{\rm pt}})$, not just the positive part, we {\it almost} recover the entire non-planar alphabet found in recent two-loop massless five-point amplitudes calculations in gauge theory \cite{Gehrmann:2015bfy,Chicherin:2017dob,Gehrmann:2018yef,Abreu:2018aqd,Chicherin:2018yne,Badger:2019djh} and gravity \cite{Chicherin:2019xeg,Abreu:2019rpt}. Interestingly, the one letter we do not obtain also seems to drop out of appropriately defined finite remainders \cite{Abreu:2018aqd,Chicherin:2018yne,Chicherin:2019xeg,Abreu:2019rpt}.

\section{The Gr\"obner fan}
\label{Grobnerfan}
The Grassmannian ${\rm Gr}(k,n)$ is the space of $k$-planes in $n$ dimensions. A point in the Grassmannian can be specified by $k$ $n$-component vectors which can be organised into a $k \times n$ matrix. These matrices are defined up to row operations which leave the plane invariant, the resulting space of matrices modulo $GL(k)$ transformations is $k(n-k)$ dimensional. 

Alternatively, the Grassmannian can be described through the set of $\binom{n}{k}$ maximal minors $p_{i_1\ldots i_k}$ or Pl\"ucker coordinates. On the set of Pl\"ucker coordinates row operations act as an overall scaling and, modulo the overall scaling, the vector of Pl\"ucker coordinates may be thought of as a point in the projective space $\mathbb{P}^{\binom{n}{k}-1}$. An arbitrary point in $\mathbb{P}^{\binom{n}{k}-1}$ is not necessarily realisable as a matrix, since the set of $k \times k$ minors of any $k \times n$ matrix are not independent, and instead obey homogenous quadratic relations known as the Pl\"ucker relations which take the form
\be
\label{Pluckerrels}
p_{i_1 \ldots i_r [ i_{r+1} \dots i_k} p_{j_1 \ldots j_{r+1}] j_{r+2} \ldots j_k}=0,
\ee
where the square brackets denote total antisymmetrisation among the $(k+1)$ indices.
We call the ideal generated by the Pl\"ucker relations inside the ring of polynomials in the Pl\"ucker coordinates the Pl\"ucker ideal $I_{k,n}$. The Grassmannian can then be thought as the projective variety inside $\mathbb{P}^{\binom{n}{k}-1}$ whose points vanish on the Pl\"ucker ideal, i.e. the vanishing set $V(I_{k,n})$. As an example consider the case of ${\rm Gr}(2,n)$, whose Pl\"ucker ideal is generated by the relations
\be
I_{2,n} = \langle p_{ij}p_{kl}-p_{ik}p_{jl}+p_{il}p_{jk}:  1 \leq i  < j  < k  < l  \leq n \rangle. 
\label{eq:G2n_gen}
\ee

Here we will actually consider the Grassmannian modulo the action of the $n$ rescalings which can be applied to the $n$ columns of the $k \times n$ matrix, so that each column actually represents an element of $\mathbb{P}^{k-1}$. Since the overall scaling is taken into account already in the action of $GL(k)$, the dimension of the space is reduced to $k(n-k)-(n-1)=(k-1)(n-k-1)$. As is common in the literature we will also refer to this space simply as the Grassmannian and henceforth when we discuss the Grassmannian we will always mean it in this reduced sense. The same scalings are present as an invariance of the Pl\"ucker relations (\ref{Pluckerrels}) which are not only homogeneous in the total degree of all variables but homogeneous in the presence of each label $i_k$ on the Pl\"ucker coordinates.

Using the general viewpoint of the Grassmannian as an ideal generated by polynomial relations we can introduce a fan structure on $\mathbb{R}^{\binom{n}{k}}$ known as the Gr\"obner fan. 
To begin to understand the structure of the Gr\"obner fan we must first introduce the notion of monomial orderings, initial ideals, and Gr\"obner bases. Our presentation follows that of \cite{boss2021grob}. 

Let $f$ be a polynomial in $n$ variables $(x_1,\ldots, x_n)$ with coefficients in an algebraically closed field $\mathbb{K}$ for which we use the notation
\be
f = \sum_{\vec{\alpha}} c_{\vec{\alpha}} {\bf x}^{\vec{\alpha}},
\ee
where we have introduced $\vec{\alpha}=(\alpha_1, \ldots, \alpha_{n}) \in \mathbb{Z}^n_{\geq 0}$, and ${\bf x}^{\vec{\alpha}}$ is understood as the monomial $x_1^{\alpha_1} \ldots x_n^{\alpha_n}$. Given some weight vector $\vec{w} \in \mathbb{R}^n$ we can define the {\it initial form} of $f$ with respect to $\vec{w}$ as 
\be
\text{in}_{\vec{w}}(f) = \sum_{\vec{\alpha}: \ \vec{\alpha}\cdot \vec{w}=m } c_{\vec{\alpha}} {\bf x}^{\vec{\alpha}},
\ee
where $m = \min\{ \vec{\alpha} \cdot \vec{w}: \ c_{\vec{\alpha}} \neq 0\}$. Furthermore, given an ideal $I \subset \mathbb{K}[x_1,\ldots, x_n]$, we can define its {\it initial ideal} with respect to $\vec{w}$ as the ideal generated by the initial forms of all functions $f \in I$ written as
\be
\text{in}_{\vec{w}}(I) = \langle \text{in}_{\vec{w}}(f): f \in I  \rangle.
\ee
If, for some finite set of generators $\mathcal{G} = \{g_1, \ldots, g_r \} \in I$, we have $\text{in}_{\vec{w}}(I) = \langle \text{in}_{\vec{w}}(g): g \in \mathcal{G}  \rangle$ we call $\mathcal{G}$ a {\it Gr\"obner basis} for $I$ with respect to $\vec{w}$. 

The next definition we need is that of a {\it monomial order}. A monomial order $<$  on the set of monomials ${\bf x}^{\vec{\alpha}} \in \mathbb{K}[x_1,\ldots,x_n]$ is a total order which satisfies \begin{align*}
&i) \ 1 \leq {\bf x}^{\vec{\alpha}} ,\\
&ii) \text{ if }  {\bf x}^{\vec{\alpha}} <  {\bf x}^{\vec{\beta}} \implies \ {\bf x}^{\vec{\alpha} + \vec{\gamma}} <  {\bf x}^{\vec{\beta} + \vec{\gamma}}.
\end{align*} 
This allows us to define the leading monomial of the polynomial $f$ as $\text{in}_<(f) = c_{\vec{\beta}}{\bf x}^{\vec{\beta}}$, where ${\bf x}^{\vec{\beta}}$ is the leading monomial with respect to $<$ appearing in $f$ with non-zero coefficient i.e. ${\bf x}^{\vec{\beta}}=\max_< \{{\bf x}^{\vec{\alpha}}: c_{\vec{\alpha}} \neq 0 \}$. Similarly, we can define the initial ideal of $I$ with respect to $<$ as
\be
\text{in}_<(I) = \langle \text{in}_<(f): f \in I  \rangle.
\ee
Note, we may always choose some weight vector $\vec{w} \in \mathbb{N}^n$ such that $\text{in}_{\vec{w}}(I) = \text{in}_{<}(I)$ \cite[Theorem 3.2.1]{HH_book}. The converse is not generally true however. 

By varying the weight vector $\vec{w}$ we may study all possible initial ideals of $I$. This leads us to the notion of the Gr\"obner fan $GF(I)$ on $\mathbb{R}^n$ as follows: two weight vectors $\vec{w}_1$ and $\vec{w}_2$ lie in the relative interior of the same cone $C$ if and only if $\text{in}_{\vec{w}_1}(I)=\text{in}_{\vec{w}_2}(I)$ i.e. they generate the same initial ideal \cite{MoraRobbiano_GF}. Note, each full-dimensional (maximal) Gr\"obner cone is associated to a monomial initial ideal specified by some monomial order $<$, consisting of all weight vectors $\vec{w}\in\mathbb{R}^n$ such that $\text{in}_{\vec{w}}(I) = \text{in}_{<}(I)$. A weight vector will lie on the boundary of a maximal cone when the associated initial ideal is no longer monomial. This collection of maximal Gröbner cones and their intersections is the Gröbner fan $GF(I)$.

Note that the Gr\"obner fan may have a linear subspace (or lineality space) consisting of elements $\vec{l}$ such that $\text{in}_{\vec{l}}(I)=I$. This is the case, for example, if the ideal is homogeneous.
We are always free to shift a weight vector by any element of the lineality space of the Gr\"obner fan $GF(I)$ without altering the initial ideal. Therefore it makes sense to consider the Gr\"obner fan modulo the action of this linear subspace. In the case of the Grassmannians ${\rm Gr}(k,n)$ we have an $n$-dimensional lineality space corresponding to the $n$ column rescalings mentioned previously. This is precisely because the Pl\"ucker relations are homogeneous with respect to the appearance of all labels $i_r$ on the Pl\"ucker variables $p_{i_1 \dots p_k}$.

We will also be interested in certain subfans of the Gr\"obner fan, the first being the tropical fan $\text{Trop}(I)$ defined as the subfan
\be
\text{Trop}(I) = \{  \vec{w} \in \mathbb{R}^n: \text{in}_{\vec{w}}(I) \text{ contains no monomial} \}.
\ee
We may restrict further and define the totally positive tropical fan $\text{Trop}^+(I)$ given by
\be
\text{Trop}^+(I) = \{  \vec{w} \in \text{Trop}(I): \text{in}_{\vec{w}}(I) \text{ is totally positive} \},
\ee 
where we an ideal $I \subset \mathbb{R}[x_1,\ldots,x_n]$ is called {\it totally positive} if it does not contain any non-zero polynomial with all positive coefficients 
(or equivalently, it has an initial ideal whose vanishing set intersects the positive orthant \cite{einsiedler}).

The above discussion is most easily demonstrated with an example, the simplest case being ${\rm Gr}(2,4)$, whose Pl\"ucker ideal is generated by a single polynomial
\be
I_{2,4} = \langle p_{12}p_{34}-p_{13}p_{24}+p_{14}p_{23} \rangle \subset \mathbb{R}[p_{12},p_{13},p_{14},p_{23},p_{24},p_{34}].
\ee
Let $\vec{w} = (w_{12},w_{13},w_{14},w_{23},w_{24},w_{34}) \in \mathbb{R}^{6}$ and $f=p_{12}p_{34}-p_{13}p_{24}+p_{14}p_{23}$ which, being the sole generator of $I_{2,4}$, constitutes a Gr\"obner basis for any choice of weight vector $\vec{w}$. A generic weight vector $\vec{w} \in \mathbb{R}^6$ can always be bought to the form $(x,y,0,\ldots,0)$ with some suitable choice of lineality shift in the four-dimensional linear subspace. The resulting Gr\"obner fan $GF(I_{2,4})$ is depicted in the $(x,y)$ plane in Figure \ref{GF(2,4)}. The Gr\"obner fan $GF(I_{2,4})$ consists of three maximal cones labelled by the monomial initial ideals $\langle p_{12}p_{34} \rangle$, $\langle p_{13}p_{24} \rangle$ and $\langle p_{14}p_{23} \rangle$. The  maximal cones intersect to give the three rays of the tropical fan $\text{Trop}(I_{2,4})$ given by $e_{12}= (1,0)$, $e_{13}=(0,1)$ and $e_{14}=(-1,-1)$, whose corresponding binomial initial ideals are $\langle p_{14}p_{23}-p_{13}p_{24} \rangle$, $\langle p_{12}p_{34}+p_{14}p_{23} \rangle$ and $\langle p_{12}p_{34}-p_{13}p_{24} \rangle$. The positive part of the tropical fan $\text{Trop}^+(I_{2,4})$ consists of the rays $e_{12}= (1,0)$ and $e_{14}=(-1,-1)$ highlighted in red whose generators contain terms of opposite sign, the ray $e_{13}=(0,1)$ is not contained in $\text{Trop}^+(I_{2,4})$ since the corresponding initial ideal is generated by a polynomial with all positive coefficients. 
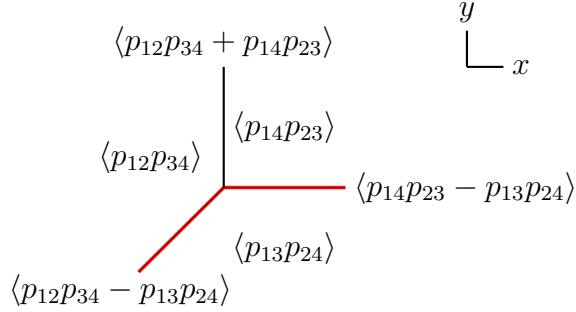
\begin{figure}[h]
\begin{center}
\begin{tikzpicture}[scale=0.8]
   \coordinate (a00) at (0,0);
   \coordinate (a10) at (2,0);
   \coordinate (a01) at (0,2);
    \coordinate (a11) at (-1.4,-1.4);
    \coordinate (la11) at (-1.7,-1.7);
     \coordinate (xy00) at (4,2);
      \coordinate (xy10) at (4.6,2);
      \coordinate (xy01) at (4,2.6);
  
    \coordinate (m1) at (1,1);
    \coordinate (m2) at (1,-1);
     \coordinate (m3) at (-1.2,0.5);
 
   \draw[thick] (a00) -- (a11);
   \draw[line width=0.5mm,opacity=0.7,red] (a00)--(a11);
   \draw[thick] (a00) -- (a10);
     \draw[line width=0.5mm,opacity=0.7,red] (a00)--(a10);
   \draw[thick] (a00) -- (a01);
   
    \node[label=above:$\langle p_{12}p_{34}+p_{14}p_{23} \rangle$] at (a01) {};
    \node[label=right:$\langle p_{14}p_{23}-p_{13}p_{24} \rangle$] at (a10) {};
     \node[] at (la11) {$\langle p_{12}p_{34}-p_{13}p_{24} \rangle$};
     
     \node[] at (m1) {$\langle p_{14}p_{23} \rangle$};
      \node[] at (m2) {$\langle p_{13}p_{24} \rangle$};
      \node[] at (m3) {$\langle p_{12}p_{34} \rangle$};
       \node[label=right:$x$] at (xy10) {};
         \node[label=above:$y$] at (xy01) {};

       \draw[thick] (xy00) -- (xy10);
       \draw[thick] (xy00) -- (xy01);

\end{tikzpicture}
\end{center}
\caption{\small The Gr\"obner fan structure of $GF(I_{2,4})$ with each region labelled by its initial ideal.  Each point in the $(x,y)$ plane corresponds to a $4$-dimensional linear subspace of $\mathbb{R}^6$ consisting of all weight vectors lineality equivelent to $(x,y,0,0,0,0)$. The tropical fan corresponds to the three rays, whilst the positive tropical fan corresponds to the two red rays.}
\label{GF(2,4)}
\end{figure}
\subsection{Polynomial cluster variables and forbidden pairs}
Moving forward, the final definition needed is that of a prime (or alternatively non-prime) ideal. An ideal $I$ is non-prime if there exists two polynomials $f \not\in I$ and $g \not\in I$ such that their product $f \cdot g \in I$. In this case we call $f$ and $g$ non-prime factors of $I$. Note, a non-prime ideal can always be decomposed into the intersection of finitely many prime components.

With all the necessary material reviewed let us remind ourselves of our goal: to extract (at least in the finite cases) the cluster variables and adjacency relations of the Grassmannian cluster algebras $\text{Gr}(k,n)$ from the Gr\"obner fan $GF(I_{k,n})$, which in the cases of $\text{Gr}(4,6)$ and $\text{Gr}(4,7)$ provide vital information for the amplitude bootstrap in the form of the symbol alphabet and adjacency rules. The ideas we will make use of have been presented in \cite{boss2021grob} for the case $\text{Gr}(2,n)$ and $\text{Gr}(3,6)$ and more generally for any cluster algebra of geometric finite type in \cite{Ilten-Najera-Treff}.

{\bf Cluster variables:} The Pl\"ucker ideal is defined on the $\binom{n}{k}$ Pl\"ucker coordinates ${p_{i_1,\ldots,i_k}}$. In the case of the Grassmannians ${\rm Gr}(2,n)$ these make up the full set of cluster variables. However, for ${\rm Gr}(3,6)$, and more importantly ${\rm Gr}(3,7)$ relevant for heptagon amplitudes, cluster variables quadratic in the Pl\"ucker coordinates start to appear. As we shall explain these {\it missing} cluster variables appear as {\it non-prime factors of initial ideals inside the maximal cones of $\text{Trop}^+(I_{k,n})$}.

{\bf Forbidden pairs:} Generally, the rays of the positive tropical fan $\text{Trop}^+(I_{k,n})$ will span multiple maximal Gr\"obner cones. That is to say taking a suitably generic\footnote{Weight vectors not lying in the intersection of maximal cones of the Gr\"obner fan.} set of weight vectors $\vec{w}$ lying in the span of the rays of $\text{Trop}^+(I_{k,n})$ we will generate multiple monomial initial ideals ${\rm in}_{\vec{w}}(I_{k,n})$. 
However, upon extending the ideal $I_{k,n}$ by the missing cluster variables to the ideal $I_{k,n}^{\text{ext}}$, the rays of $\text{Trop}^+(I^{\text{ext}}_{k,n})$ span a single maximal Gr\"obner cone in $GF(I^{\text{ext}}_{k,n})$. The minimal generating set of the initial ideal of this maximal Gr\"obner cone provides us with a list of monomials which are exactly the forbidden pairs of cluster variables. Note, in the case of ${\rm Gr}(2,n)$ all cluster variables are present already for $I_{2,n}$ and no extension procedure is needed. The rays of $\text{Trop}^+(I_{2,n})$ already span a unique maximal Gr\"obner cone.

In the remainder of this section we review these ideas in more detail for the case of ${\rm Gr}(2,5)$. Later in Section \ref{sec:Adjacency} we return to the case of ${\rm Gr}(3,6)$, discussed in \cite{boss2021grob}, and further apply this discussion to ${\rm Gr}(3,7) \cong {\rm Gr}(4,7)$ relevant for heptagon amplitudes. We also discuss the remaining finite case ${\rm Gr}(3,8)$ and the outlook beyond the finite type Grassmannians. Finally, in Section \ref{spinor} we will apply some of the ideas to a non-Grassmannian case relevant for scattering in more general massless theories.
\subsection{$GF(I_{2,5})$}
\label{GF25}

We conclude this section with the example of ${\rm Gr}(2,5)$. In the space of Pl\"ucker coordinates $( p_{12}, \ldots ,p_{45} )$ the five Pl\"ucker relations are given by 
\be
p_{ij}p_{kl}-p_{ik}p_{jl}+p_{il}p_{jk}=0, \quad 1 \leq i  < j  < k  < l  \leq 5.
\label{eq:G25_gen}
\ee
The Gr\"obner fan $GF(I_{2,5})$ is simplicial, containing $132$ maximal cones and twenty rays. Arranging the coordinates in lexicographic order 
\be
\{w_{12},w_{13},w_{14},w_{15},w_{23},w_{24},w_{25},w_{34},w_{35},w_{45}\},
\ee
the rays are defined as
\begin{align}
e_{12} &= (1,0,0,0,0,0,0,0,0,0)\,,\notag \\
&\,\,\, \vdots\notag\\
e_{45} &= (0,0,0,0,0,0,0,0,0,1),
\label{G25rays}
\end{align}
along with ten more given by $-e_{ij}$. 

\begin{figure}[h]
\begin{center}
  \begin{tikzpicture}[label distance = 0.18]
    \pgfmathsetmacro{\innerrad}{1.8};
    
    \coordinate (i1) at (90:\innerrad);
    \coordinate (i2) at (234:\innerrad);
    \coordinate (i3) at (18:\innerrad);
    \coordinate (i4) at (162:\innerrad);
    \coordinate (i5) at (-54:\innerrad);

    \coordinate (o1) at (90:3);
    \coordinate (o2) at (234:3);
    \coordinate (o3) at (18:3);
    \coordinate (o4) at (162:3);
    \coordinate (o5) at (-54:3);

    \node[label=above left:$e_{12}$] at (i1) {};
    \node[label=left:$e_{45}$] at (i2) {};
    \node[label=above:$e_{23}$] at (i3) {};
    \node[label=above:$e_{15}$] at (i4) {};
    \node[label=right:$e_{34}$] at (i5) {};

    \node[] at ($(o1)+0.12*(o1)$) {$e_{35}$};
    \node[] at ($(o2)+0.12*(o2)$) {$e_{13}$};
    \node[] at ($(o3)+0.12*(o3)$) {$e_{14}$};
    \node[] at ($(o4)+0.12*(o4)$) {$e_{24}$};
    \node[] at ($(o5)+0.12*(o5)$) {$e_{25}$};

    \draw[line width=0.5mm,opacity=1,red] (i1) -- (i2) -- (i3) -- (i4) -- (i5) -- cycle;
    \draw[thick,black] (i1) -- (i2) -- (i3) -- (i4) -- (i5) -- cycle;
    \draw (o1) -- (o3) -- (o5) -- (o2) -- (o4) -- cycle;
    \foreach \i in {1,...,5}{
      \draw (i\i) -- (o\i);
    }
    \foreach \i in {1,...,5}{
      \draw[fill=red] (i\i) circle (0.07);
      \draw[fill=black] (o\i) circle (0.07);
    }

  \end{tikzpicture}
\caption{\small The 10 vertices and 15 edges of ${\rm Trop}(I_{2,5})$ space. The positive part is highlighted red.}
\label{Petersen_graph}
\end{center}
\end{figure}
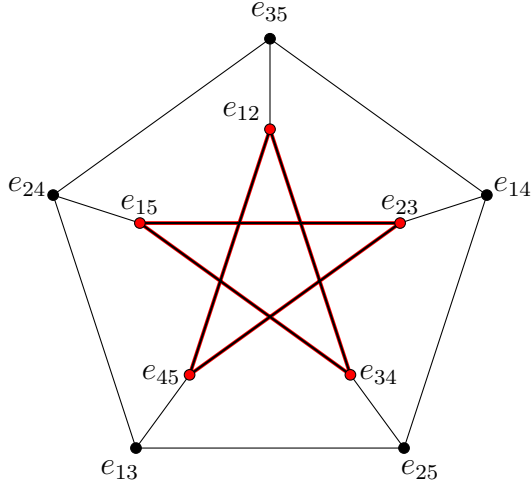

The Gr\"obner fan contains twelve maximal cones with five rays given by $e_{ij}$ from the list (\ref{G25rays}). A further 60 maximal cones have four of the $e_{ij}$ and one of the $-e_{ij}$ rays and $60$ more maximal cones have three $e_{ij}$ and two $-e_{ij}$ rays. The ten $e_{ij}$ vectors defined in (\ref{G25rays}) make up the rays of the tropical fan ${\rm Trop}(I_{2,5})$. They are connected in a Petersen graph topology shown in Figure \ref{Petersen_graph}. The positive tropical fan ${\rm Trop}^+(I_{2,5})$ has five rays given by
\be
\{e_{12},e_{23},e_{34},e_{45},e_{15} \}
\label{trop+25rays}
\ee 
which are highlighted in red.

There is exactly one maximal Gr\"obner cone
spanned by the five rays (\ref{trop+25rays}) of $\text{Trop}^+(I_{k,n})$. This cone has a Gr\"obner basis whose initial monomials are all the crossing chords of the pentagon, i.e. 
\be
{\rm in}_{\vec{w}}(I_{2,5}) =\langle p_{13}p_{24},p_{13}p_{25},p_{14}p_{25},p_{14}p_{35},p_{24}p_{35}\rangle\,
\ee 
for any $\vec{w}$ given by a strictly positive linear combination of the five rays of ${\rm Trop}^{+}(I_{2,5})$.

The case of ${\rm Gr}(2,n)$ was studied in detail in \cite{boss2021grob}, where a similar construction was shown to hold for all $n$. The rays of the positive tropical fan ${\rm Trop}^+(I_{2,n})$ span a single maximal Gr\"obner cone inside $GF(I_{2,n})$, whose initial ideal is generated products of Pl\"ucker variables labelled by crossing chords of the $n$-gon. Note that this includes the case ${\rm Gr}(2,6) \cong {\rm Gr}(4,6)$ relevant for hexagon amplitudes in planar $\mathcal{N}=4$ super Yang-Mills theory. As was also shown in \cite{boss2021grob} for the case of ${\rm Gr}(3,6)$, and as we shall explain in Section \ref{sec:Adjacency}, it is not the case for $k>2$ that the rays of ${\rm Trop}^+(I_{k,n})$ span a single maximal Gr\"obner cone inside $GF(I_{k,n})$. Instead, to identify a single maximal Gr\"obner cone, whose rays are given by the positive tropical part, the ideal must first be extended by the missing cluster variables which are polynomials in Pl\"ucker variables.
\section{$\text{Trop}^+(I_{k,n})$ and the Speyer--Williams fan}
\label{sec:cluster}
In practice, it is only possible to compute the entire Gr\"obner fan for the most simple of cases. Therefore, it is highly desirable to have an efficient route to calculating the positive tropical part $\text{Trop}^+(I_{k,n})$ directly without having it embedded in the Gr\"obner fan, or even in ${\rm Trop}(I_{k,n})$. In this section we review the methods of \cite{speyer2003tropical} on this direct construction. For further details of the structure of the resulting fans see \cite{Drummond:2019qjk,Drummond:2020kqg}. The cases of interest will be $\text{Gr}(3,n)$ for $n=6,7,8$  which, along with $\text{Gr}(2,n)$, make up the finite type Grassmannian cluster algebras.

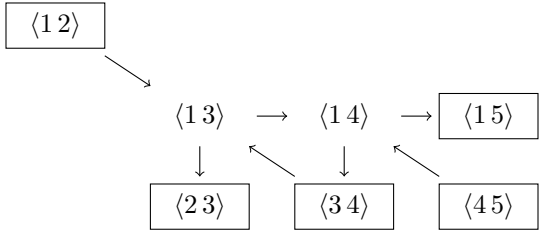
\begin{figure}[h]
\begin{center}
{\footnotesize
\begin{tikzpicture}
\pgfmathsetmacro{\nw}{1.3}
\pgfmathsetmacro{\vvwnw}{2.5}
\pgfmathsetmacro{\vvvwnw}{2.85}
\pgfmathsetmacro{\nh}{0.6}
\pgfmathsetmacro{\aa}{0.6}
\pgfmathsetmacro{\ep}{0.1}
\node at (-0.5*\nw -\aa,\aa+0.5*\nh) {$\langle 1\,2 \rangle$};
\draw[] (-\aa,\aa) -- (-\aa -\nw,\aa) -- (-\aa -\nw, \aa+\nh) -- (-\aa,\aa+\nh) -- cycle;
\node at (0.5*\nw +0*\aa,-0*\aa-0.5*\nh) {$\langle 1\,3 \rangle$};
\node at (0.5*\nw +0*\aa,-1*\aa-1.5*\nh) {$\langle 2\,3 \rangle$};
\node at (1.5*\nw +1*\aa,-0*\aa-0.5*\nh) {$\langle 1\,4 \rangle$};
\node at (1.5*\nw +1*\aa,-1*\aa-1.5*\nh) {$\langle 3\,4 \rangle$};

\node at (2.5*\nw +2*\aa,-0*\aa-0.5*\nh) {$\langle 1\,5 \rangle$};
\draw[] (2*\nw+2*\aa,-0*\aa-0*\nh) -- (2*\nw+2*\aa,-0*\aa-1*\nh) -- (3*\nw+2*\aa,-0*\aa-1*\nh) -- (3*\nw+2*\aa,-0*\aa-0*\nh) -- cycle;
\node at (2.5*\nw +2*\aa,-1*\aa-1.5*\nh) {$\langle 4\,5 \rangle$};

\draw[] (2*\nw+2*\aa,-1*\aa-1*\nh) -- (2*\nw+2*\aa,-1*\aa-2*\nh) -- (3*\nw+2*\aa,-1*\aa-2*\nh) -- (3*\nw+2*\aa,-1*\aa-1*\nh) -- cycle;
\draw[] (1*\nw+1*\aa,-1*\aa-1*\nh) -- (1*\nw+1*\aa,-1*\aa-2*\nh) -- (2*\nw+1*\aa,-1*\aa-2*\nh) -- (2*\nw+1*\aa,-1*\aa-1*\nh) -- cycle;
\draw[] (0*\nw+0*\aa,-1*\aa-1*\nh) -- (0*\nw+0*\aa,-1*\aa-2*\nh) -- (1*\nw+0*\aa,-1*\aa-2*\nh) -- (1*\nw+0*\aa,-1*\aa-1*\nh) -- cycle;

\draw[->] (-\aa+0*\ep,\aa-\ep) -- (0-0*\ep,0+\ep);
\draw[->] (0.5*\nw,-\nh-\ep) -- (0.5*\nw,-\nh-\aa+\ep);
\draw[->] (1*\nw+\ep,-0.5*\nh) -- (1*\nw+\aa-\ep,-0.5*\nh);
\draw[->] (1*\nw+\aa-0*\ep,-\nh-\aa+\ep) -- (1*\nw+0*\ep,-\nh-\ep);
\draw[->] (1.5*\nw+1*\aa,-\nh-\ep) -- (1.5*\nw+1*\aa,-\nh-\aa+\ep);
\draw[->] (2*\nw+\aa+\ep,-0.5*\nh) -- (2*\nw+2*\aa-\ep,-0.5*\nh);
\draw[->] (2*\nw+2*\aa-0*\ep,-\nh-\aa+\ep) -- (2*\nw+\aa+0*\ep,-\nh-\ep);
\end{tikzpicture}
}
\end{center}
\caption{\small The initial cluster of ${\rm Gr}(2,5)$.
}
\label{Gr25initial}
\end{figure}

Generally, the ${\rm Gr}(k,n)$ initial cluster has the form of a $(k-1)\times(n-k-1)$ array of active (mutable) nodes, in addition to $k$ frozen nodes, each labelled by an $\mathcal{A}$-coordinate. An example of the initial cluster for the case of ${\rm Gr}(2,5)$ is given in Figure \ref{Gr25initial}, with frozen nodes indicated as the boxed vertices. We may also assign to each active node an $\mathcal{X}$-coordinate, given by the product of incoming $\mathcal{A}$-coordinates over the product of the outgoing ones. Following the array of active nodes, these also organise themselves into a $(k-1)\times(n-k-1)$ array $X$ with elements $x_{rs}$. 

Using the $x_{rs}$ we can define the $k\times n$ web matrix $W^{(k,n)}=({\bf I}_k|M)$. Where $M$ is given by the $k \times (n-k)$ matrix elements
\begin{align}
\label{Genwebmatrix}
m_{ij}= (-1)^{i+k} \sum_{\vec{\lambda} \in Y_{ij}} \prod_{r=1}^{k-i} \prod_{s=1}^{\lambda_{r}}x_{rs},
\end{align}
with the summation range $Y_{ij}$ given by $0 \leq \lambda_{k-i} \leq ... \leq \lambda_1 \leq j-1$. The web matrix thus allows us to evaluate all $\mathcal{A}$-coordinates as subtraction free polynomials in the $\mathcal{X}$-coordinates by identifying the Pl\"ucker coordinates with the maximal minors of the web matrix i.e.
\be
p_{i_i \ldots i_k}= \det(W^{(k,n)}_{i_1,\ldots, i_k})(x_{rs})
\ee
 where $W^{(k,n)}_{i_1,\ldots, i_k}$is understood as the matrix formed by taking columns $i_1 \ldots i_k$ of $W^{(k,n)}$.

Let us illustrate this with the example of ${\rm Gr}(2,5)$. The initial cluster is depicted in Figure \ref{Gr25initial} from which we can read off the $\mathcal{X}$-coordinates, they are given by  
\begin{align}
x_{11} &= \frac{\langle 12 \rangle \langle 34 \rangle }{\langle 14 \rangle \langle 23 \rangle}\,, \quad &&x_{12} = \frac{\langle 13 \rangle \langle 45 \rangle}{\langle 34 \rangle \langle 15 \rangle} \,.
\end{align}
With the $\mathcal{X}$-coordinates to hand we can write down the web matrix,
\be
{\small
W^{(2,5)}\!=\!\left[
\begin{matrix}
1 &\, 0 &\, -1 &\, -1-x_{11} &\, -1 -x_{11}-x_{11}x_{12}\\
0 &\, 1 &\, 1 &\,1 & \, 1
\end{matrix}
\right].
}
\label{webmatrix}
\ee
By identifying the Pl\"ucker coordinates $p_{ij}$ with the maximal minor formed by columns $i$ and $j$ of the web matrix as
\be
p_{ij} = \det (W^{(2,5)}_{ij})(x_{11},x_{12}),
\ee
we immediately arrive at an expression for all $\mathcal{A}$-coordinates as subtraction free polynomials in the $\mathcal{X}$-coordinates. As an example we have
$$
p_{25} = x_{11} + x_{11} x_{12},
$$
We can now consider tropicalising the expressions for the $\mathcal{A}$-coordinates, which amounts to replacing $(+,\times)$  by their tropical counterparts $(\text{min},+)$. The tropical version of the above expression for $p_{25}$ is given by
\be
\tilde{p}_{25} = {\rm min}(\tilde{x}_{11},\tilde{x}_{11}+\tilde{x}_{12})\,,
\ee 
where $\tilde{p}$ and $\tilde{x}$ are used to emphasise that we are dealing with tropical expressions. This tropical expression defines a piecewise linear map on $\mathbb{R}^2$ with coordinates $\{ \tilde{x}_{11},\tilde{x}_{12}\}$, where regions of linearity are separated by tropical hypersurfaces, and as such provide a fan structure on the space of $\{ \tilde{x}_{11}, \tilde{x}_{12}\}$. For example the tropical hypersurface of $\tilde{p}_{25}$ is given by
\begin{align}
\tilde{x}_{12} = 0\,.
\end{align}
By tropicalising different subsets of the $\mathcal{A}$-coordinates we can define different tropical fans given by the common refinement of all fans in the subset of tropical expressions. In practice, we calculate the refinement of the tropical fan for a subset, $\mathcal{S}$, of $\mathcal{A}$-coordinates via the Minkowski sum of their Newton polytopes. The tropical expressions for the frozen coordinates do not contain any tropical hypersurface and hence do not contribute to the structure of the fan. 

Our focus will be on two fans in particular: the Speyer--Williams fan \cite{speyer2003tropical}, which is obtained by tropicalising  the set of all Pl\"ucker coordinates; and the cluster fan, where we choose to tropicalise the entire set of $\mathcal{A}$-coordinates. Note, for the case of ${\rm Gr}(2,n)$ the Pl\"ucker coordinates do make up the entire set of $\mathcal{A}$-coordinates and hence the Speyer--Williams and cluster fans coincide. However, when considering ${\rm Gr}(3,6)$ $\mathcal{A}$-coordinates quadratic in the Pl\"ucker variables begin to appear and hence the structure of the two fans begins to differ.

While we have used the initial cluster to explain the definition of the $x_{rs}$ we would like to emphasise that we are merely using them as a convenient choice of coordinates. We will not be using any of the mutation structure of the cluster algebra and indeed one of the main points of this investigation is to recover information about the Grassmannian cluster algebras (including the set of $\mathcal{A}$-coordinates) just by carefully studying the Gr\"obner structure of the Pl\"ucker ideals.
\subsection{${\rm Gr}(2,5)$}
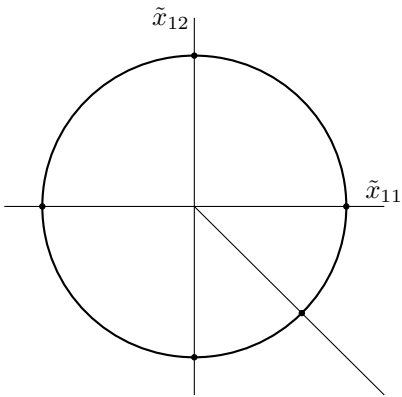
\begin{figure}[H]
	\centering
		{\footnotesize
                   \begin{tikzpicture}[node distance=2cm, scale=0.5]
      \draw[opacity=0] (-5.2,-5,2) rectangle (5.2,5.2);
    \draw[] (-5,0) --(5,0) node[above]{$\tilde{x}_{11}$};
    \draw[] (0,-5) --(0,5) node[left]{$\tilde{x}_{12}$};
    \draw[] (0,0) --(5,-5) ;
    \draw[thick] (0,0) circle (4);
    \foreach \ang in {0, 90, 180, 270, 315}{
      \draw[fill=black] (\ang:4) circle (0.07);
    }
  \end{tikzpicture}}
              \caption{\small The tropical fan $\text{Trop}^+(I_{2,5})$ via the Speyer--Williams construction.}
              \label{fan25}
\end{figure}

 The ${\rm Gr}(2,5)$ web matrix, written in \eqref{webmatrix}, allows us to write the $10$ Pl\"ucker variables $p_{ij}$ in terms of the two $\mathcal{X}$-coordinates $(x_{11},x_{12})$ as  
\begin{align}
p_{1i} &= p_{23} = 1, & p_{24} &= 1+x_{11}, & p_{25} &= 1+x_{11}+x_{11} x_{12}, \notag \\
p_{34} &= x_{11}, & p_{35} &= x_{11} +x_{11}x_{12}, & p_{45} &= x_{11} x_{12}.
\end{align}
Taking these expressions we can compute the corresponding tropical fan via the Minkowski sum operation of e.g. {\tt gfan} \cite{gfan} and obtain the Speyer--Williams fan. The resulting fan depicted in Figure \ref{fan25}. It has five regions of linearity whose boundaries are given by the five rays
\begin{equation}
\{ (1,0), (0,1), (-1,0),(0,-1),(1,-1)\}.
\end{equation}
Note, the tropical fan as described is parameterised in the space of the $(\tilde{x}_{11},\tilde{x}_{12})$ variables. However, as discussed in \cite{Drummond:2020kqg,Drummond:2019qjk}, we can map the five rays above to those of the positive tropical Grassmannian \cite{SpeyerSturmfels} presented in Section \ref{GF25}, by taking the scalar product of the unit vectors $e_{ij}$ with the vector of tropicalised Pl\"ucker coordinates evaluated at particular values of the $\tilde{x}$ variables, i.e.
\be
\label{evmap25}
{\rm \bf ev}: (\tilde{x}_{11},\tilde{x}_{12}) \mapsto \sum_{1\leq i < j \leq 5}  \tilde{p}_{ij}(\tilde{x}_{11},\tilde{x}_{12}) e_{ij}.
\ee
The five rays given above up to lineality map to $\{ e_{12}, e_{45},e_{23},e_{15},e_{34}\}$. In particular, the regions between the rays of Figure \ref{fan25} map to the red edges between the corresponding rays in Figure \ref{Petersen_graph}.
The positive tropical Grassmannian ${\rm Trop}^+(I_{2,5})$ is the image under the map (\ref{evmap25}) of the Speyer--Williams fan. In particular, note that in this case the fan structures map precisely. In fact this is the case for all ${\rm Gr}(2,n)$ Grassmannians, but for higher Grassmannians, e.g. ${\rm Gr}(3,6)$ the fan structure of ${\rm Trop}^+(I_{k,n})$ is a refinement of that inherited from the Speyer--Williams fan.

\section{Alphabets and adjacency from the Gr\"obner fan}
\label{sec:Adjacency}
In the case of ${\rm Gr}(2,n)$, the Speyer--Williams fan and the cluster fan coincide since the full set of $\mathcal{A}$-coordinates are given solely by Pl\"uckers. For the remaining finite-type Grassmannians this is no longer the case. In this section we wish to begin with the Speyer--Williams fan, by tropicalising only the Pl\"ucker coordinates, and see where the additional information of the missing $\mathcal{A}$-coordinates and adjacency conditions is hidden inside the structure of $\text{Trop}^+(I_{k,n})$ and $GF(I_{k,n})$. This question is inspired by the ideas appearing in \cite{boss2021grob}. Before we begin let us remind ourselves where we will find this additional information:

{\bf Cluster variables}: The missing $\mathcal{A}$-coordinates appear as non-prime factors in the initial ideals of maximal cones of $\text{Trop}^+(I_{k,n})$.

{\bf Forbidden pairs}: Upon extending the ideal by the missing $\mathcal{A}$-coordinates the rays of $\text{Trop}^+(I_{k,n})$ span a single maximal Gr\"obner cone. The initial ideal of this maximal Gr\"obner cone, of the extended ideal, provides us with a list of monomials which are exactly the forbidden pairs of $\mathcal{A}$-coordinates.
\subsection{${\rm Gr}(3,6)$}\label{sec:Gr36}
The case of ${\rm Gr}(3,6)$ was covered in detail in \cite{boss2021grob} and we review here the relevant parts of the discussion. The Pl\"ucker Ideal $I_{3,6}$ is generated by three and four-terms relations of the form
\begin{align}
p_{123}p_{145}+p_{125}p_{134}-p_{124}p_{135}& =0, \ldots\\
p_{123}p_{456}-p_{156}p_{234}+p_{146}p_{235}-p_{145}p_{236}&=0, \ldots
\end{align}
in the ring of polynomials in the $20$ Pl\"ucker coordinates $p_{ijk}$. 

The tropical fan $\text{Trop}(I_{3,6})$ was studied in \cite{SpeyerSturmfels,HJJS,Jensenwebpage36}. It is simplicial with an $f$-vector given by
\be
(1,65,550,1395,1035),
\label{fvectTropGr36}
\ee
The $65$ rays are given by the permutation copies of the following three basic types\cite{SpeyerSturmfels,Cachazo:2019ngv},
  \begin{align}
      &e_{123}, \notag\\
      &e_{123}+ e_{124} + e_{134} + e_{234}, \notag \\
      &e_{123}+ e_{124} + e_{134} + e_{234} + e_{125} + e_{126}\,.
      \label{Trop36rays}
\end{align}
The positive part of the tropical fan $\text{Trop}^+(I_{3,6})$ is spanned by $16$ rays given by the cyclic copies of the rays listed above in (\ref{Trop36rays}) (six of the first type, six of the second and four of the third).
By restricting to the subfan spanned by these rays we obtain the positive tropical fan $\text{Trop}^+(I_{3,6})$ with the $f$-vector
\be
(1,16,68,104,52).
\ee

To obtain the corresponding Speyer--Williams fan we first evaluate all Pl\"ucker coordinates as minors of the web matrix $W^{(3,6)}$, as defined in (\ref{Genwebmatrix}). Then we tropicalise the resulting polynomials as explained in Section \ref{sec:cluster}. Note, the {\it frozen} Pl\"ucker coordinates $p_{ii+1i+2}$ are themselves monomial and do not contribute to the structure of the fan. The fan obtained is spanned by $16$ rays given by \cite{speyer2003tropical}
\begin{align}
&(1,0,0,0), &&(-1,0,0,0), &&(1,-1,0,0), &&(0,0,1,-1), \notag \\
&(0,1,0,0), &&(0,-1,0,0), &&(1,0,-1,0), &&(-1,0,0,1), \notag \\
&(0,0,1,0), &&(0,0,-1,0), &&(1,0,0,-1), &&(0,1,1,-1),\notag \\
&(0,0,0,1), &&(0,0,0,-1), &&(0,1,0,-1), &&(1,-1,-1,0)  \,,
\label{eq:G36_rays}
\end{align}
in $\tilde{x}$ space with the ordering $(\tilde{x}_{11},\tilde{x}_{21},\tilde{x}_{12},\tilde{x}_{22})$. The maximal cones of the fan are four-dimensional regions within which all minors are linear, which can be intersected with the unit sphere to produce $3$-dimensional facets of a polyhedral complex. The fan has $48$ maximal facets given by $46$ terahedra and $2$ bipyramids. They themselves have $2$-dimensional boundaries corresponding to some minor being between two regions of linearity. There are $98$ of these $2$-dimensional boundaries, which themselves are bounded by $66$ edges, which are further bounded by $16$ points. The $16$ points correspond to the intersection of the rays in \eqref{eq:G36_rays} with the unit sphere. This information can be summarised by the $f$-vector given by $f_{3,6}=(1,16,66,98,48)$. Sometimes we would also like to keep information on the number of vertices of each facet, for this we use the notation
$$
 f_{3,6}=(1_0,16_1,66_2,98_3,46_4 + 2_5),
$$
where we understand the right most element as $46$ tetrahedrons ($4$-vertex objects) and two, non-simplicial, bipyramids ($5$-vertex objects). 
The two bipyramids will play a role in the following discussion so let us discuss their structure in more detail. They are each spanned by $5$ rays from (\ref{eq:G36_rays}) as follows,
\begin{align}
b_1 &= \text{span} \{ (-1,0,0,1),(0,0,1,0),(-1,0,0,0),(0,1,0,0),(0,1,1,-1) \} \notag \\
&:= \text{span} \{ b_{11}, b_{12}, b_{13}, b_{14}, b_{15} \}, \notag \\
b_2 &= \text{span}\{(1,-1,-1,0) ,(1,0,-1,0),(1,-1,0,0),(0,0,0,-1),(1,0,0,-1) \} \notag \\
&:= \text{span} \{ b_{21}, b_{22}, b_{23}, b_{24}, b_{25} \}.
\label{eq:bi_rays}
\end{align}

As discussed before in the example of ${\rm Gr}(2,5)$, we can map the Speyer--Williams fan to the positive tropical Grassmannian discussed in Section \ref{Grobnerfan} via the map
\be
{\rm \bf ev}: (\tilde{x}_{11},\tilde{x}_{12},\tilde{x}_{21},\tilde{x}_{22}) \mapsto \sum_{1\leq i < j < k \leq 6}  \tilde{p}_{ijk}(\tilde{x}_{11},\tilde{x}_{12},\tilde{x}_{21},\tilde{x}_{22}) e_{ijk},
\label{evmap36}
\ee
with the $e_{ijk}$ comprising the unit vectors in $\mathbb{R}^{20}$,
\begin{align}
e_{123} &= (1,0,\ldots,0)\,,\notag \\
&\,\,\, \vdots\notag\\
e_{456} &= (0,\ldots,0,1)\,.
\label{G36rays}
\end{align}
For example, modulo the lineality shift of $I_{3,6}$, the five rays of the bipyramid $b_1$ map to (as discussed in \cite{Drummond:2019qjk})
\begin{align}
b_{11} &\mapsto e_{123} + e_{124} + e_{125} + e_{126} + e_{134} + e_{234}\,, \notag \\
b_{12} &\mapsto  e_{123} + e_{124} + e_{134} + e_{234}\,, \notag \\
b_{13} &\mapsto e_{125} + e_{126} + e_{156} + e_{256} \,, \notag \\
b_{14} &\mapsto e_{345} + e_{346} + e_{356} + e_{456} \,, \notag \\
b_{15} &\mapsto e_{123} + e_{124} + e_{134} + e_{234} + e_{345} + e_{346}\,.
\end{align}
In particular note that image of the centre of the bipyramid can be represented in two equivalent ways (as always modulo lineality),
\be
\label{bipyramidcentre}
{\rm \bf ev} (b_{11}) + {\rm \bf ev}(b_{15}) = {\rm \bf ev}(b_{12}) + {\rm \bf ev}(b_{13}) + {\rm \bf ev}(b_{14})\,.
\ee

\begin{figure}[h]

\centering

\begin{tikzpicture}[scale=1.2]

   \draw[thick] (3.25,0.1) node[right] {{\tiny $ = $}};

    \draw[fill=black] (1.8,2) circle (0.07);

     \draw[fill=black] (1.8,-2) circle (0.07);

\draw[fill=black] ($(1.8,0)+(-0.9,0.1)$) node[left] {{\tiny $b_{12}$}\,}  circle (0.07);

       \draw[fill=black] ($(1.8,0)+(+0.9,0.1)$) node[right] {{\,\tiny $b_{14}$}}  circle (0.07);

        \draw[fill=black] ($(1.8,0)+(-0.3,-0.3)$) circle (0.07);

 \node[] at (1.8,2.2) {{\tiny $b_{11}$}};

    \draw[thick] (1.8,2)  -- ($(1.8,0)+(0.9,0.1)$);

    \draw[thick] (1.8,-2)  -- ($(1.8,0)+(0.9,0.1)$); 
 \node[] at (1.8,-2.2) {{\tiny $b_{15}$}};

    \draw[thick] (1.8,2) -- ($(1.8,0)+(-0.9,0.1)$);

    \draw[thick] (1.8,2) -- ($(1.8,0)+(-0.3,-0.3)$);

    \draw[thick] ($(1.8,0)+(-0.3,-0.3)$) --($(1.8,0)+(0.9,0.1)$);

    \draw[thick] (1.8,-2) -- ($(1.8,0)+(-0.3,-0.3)$);

    \draw[thick] ($(1.8,0)+(-0.9,0.1)$) -- ($(1.8,0)+(-0.3,-0.3)$);

    \draw[dashed] ($(1.8,2)$) -- ($(1.8,-2)$);


    \draw[dashed] ($(1.8,0)+(0.9,0.1)$) -- ($(1.8,0)+(-0.9,0.1)$);

    \draw[thick] (1.8,-2)  -- ($(1.8,0)+(0.9,0.1)$);

    \draw[thick] (1.8,-2) -- ($(1.8,0)+(-0.9,0.1)$);

    \node[] at (1.33,-0.4) {\tiny $ b_{13}$};


    \draw[thick] (5.9,0.1) node[right] {{\tiny $ + $}};

    \draw[fill=black] (5,2) circle (0.07);

     \draw[fill=black] (5,-2) circle (0.07);

       \draw[fill=black] ($(5,0)+(-0.9,0.1)$) circle (0.07);

        \draw[fill=black] ($(5,0)+(-0.3,-0.3)$) circle (0.07);

    \draw[thick] (5,2)  -- ($(5,0)+(-0.9,0.1)$);

    \draw[thick] (5,-2)  -- ($(5,0)+(-0.9,0.1)$); 
 \node[] at (5,-2.2) {{\tiny $p_{126} p_{345} - p_{125} p_{346}$}};
 
    \draw[thick] (5,2) -- ($(5,0)+(-0.9,0.1)$);

    \draw[thick] (5,2) -- ($(5,0)+(-0.3,-0.3)$);

    \draw[thick] (5,-2) -- ($(5,0)+(-0.3,-0.3)$);

    \draw[thick] ($(5,0)+(-0.9,0.1)$) -- ($(5,0)+(-0.3,-0.3)$);

    \draw[thick] ($(5,2)$) -- ($(5,-2)$);

    \draw[thick] (5,-2) -- ($(5,0)+(-0.9,0.1)$);


    \draw[thick] (9.5,0.1) node[right] {{\tiny $ + $}};

    \draw[fill=black] (8,2) circle (0.07);

     \draw[fill=black] (8,-2) circle (0.07);

 \draw[fill=black] ($(8,0)+(-0.9,0.1)$) circle (0.07);

       \draw[fill=black] ($(8,0)+(+0.9,0.1)$) circle (0.07);

    \draw[thick] (8,2)  -- ($(8,0)+(0.9,0.1)$);

    \draw[thick] (8,-2)  -- ($(8,0)+(0.9,0.1)$); 
 \node[] at (8,-2.2) {{\tiny $ p_{124} p_{356} - p_{123} p_{456} $}};

    \draw[thick] (8,2) -- ($(8,0)+(-0.9,0.1)$);

    \draw[thick] ($(8,2)$) -- ($(8,-2)$);

    \draw[dashed] ($(8,0)+(0.9,0.1)$) -- ($(8,0)+(-0.9,0.1)$);

    \draw[thick] (8,-2)  -- ($(8,0)+(0.9,0.1)$);

    \draw[thick] (8,-2) -- ($(8,0)+(-0.9,0.1)$);


    \draw[fill=black] (11,2) circle (0.07);

     \draw[fill=black] (11,-2) circle (0.07);

       \draw[fill=black] ($(11,0)+(+0.9,0.1)$) circle (0.07);

        \draw[fill=black] ($(11,0)+(-0.3,-0.3)$) circle (0.07);

    \draw[thick] (11,2)  -- ($(11,0)+(0.9,0.1)$);

    \draw[thick] (11,-2)  -- ($(11,0)+(0.9,0.1)$); 
 \node[] at (11,-2.2) {{\tiny $p_{256} p_{134} - p_{234} p_{156}$}};

    \draw[thick] (11,2) -- ($(11,0)+(-0.3,-0.3)$);

    \draw[thick] ($(11,0)+(-0.3,-0.3)$) --($(11,0)+(0.9,0.1)$);

    \draw[thick] (11,-2) -- ($(11,0)+(-0.3,-0.3)$);

    \draw[dashed] ($(11,2)$) -- ($(11,-2)$);

    \draw[thick] (11,-2)  -- ($(11,0)+(0.9,0.1)$);

\end{tikzpicture}

\caption{\small The image of the bipyramid $b_1$ inside $\text{Trop}^+(I_{3,6})$, on the left hand we have the full bipyramid with its $5$ rays, on the right the bipyramid is split into three tetrahedra by the structure of the Gr\"obner fan. Each tetrahedron is labelled by the quadratic non-prime factor found in the initial ideal. Note all three expressions are equivalent modulo the Pl\"ucker relations.}

\label{fig:grob_split_36}

\end{figure}
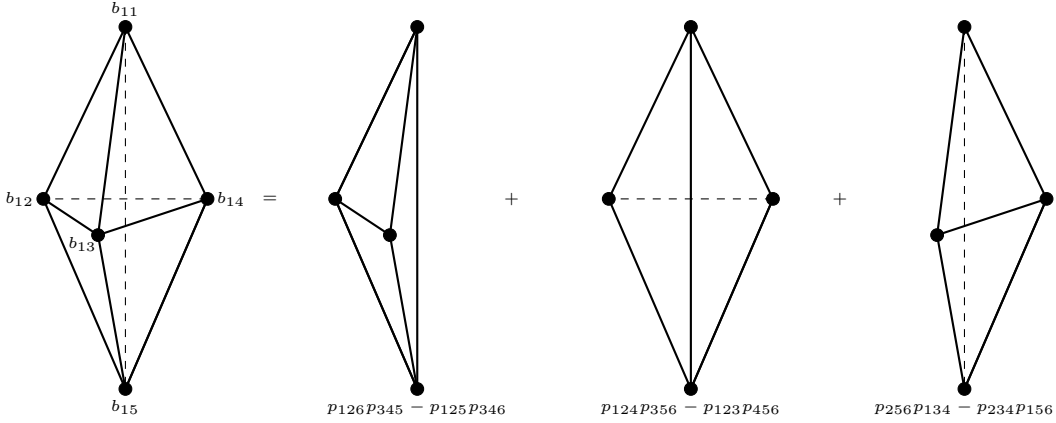

The sixteen rays given in (\ref{eq:G36_rays}) map to the sixteen rays of ${\rm Trop}^+(I_{3,6})$ but the fan structures differ slightly. In fact, ${\rm Trop}^+(I_{3,6})$ which, as introduced in Section \ref{Grobnerfan}, inherits its fan structure from the Gr\"obner fan $GF(I_{3,6})$, has two additional edges and six additional triangles in comparison to the Speyer--Williams fan. In other words, as a fan ${\rm Trop}^+(I_{3,6})$ is a refinement of (the image under ${\rm \bf ev}$ of) the Speyer--Williams fan. The additional edges and triangles actually slice each of the two bipyramids into three tetrahedra, each with its own initial ideal, as illustrated in Figure \ref{fig:grob_split_36}.

Related to the splitting of the bipyramids (and in contrast with the ${\rm Gr}(2,n)$ case), the $16$ rays of $\text{Trop}^+({I_{3,6}})$ do not single out an individual maximal Gr\"obner cone of $GF(I_{3,6})$. In fact, calculating initial ideals inside the span of the $16$ positive rays of $\text{Trop}^+({I_{3,6}})$ we find nine maximal Gr\"obner cones. The splitting of the bipyramids can be understood as each bipyramid intersecting three of the nine maximal Gr\"obner cones spanned by the rays of $\text{Trop}^+(I_{3,6})$. 

As shown in \cite{boss2021grob}, we can resolve the span of the positive tropical rays into a single maximal Gr\"obner cone by extending the ideal. To decide how to extend the ideal we search the maximal cones of $\text{Trop}^+(I_{3,6})$ for initial ideals which are not prime and whose factors will provide for us the missing $\mathcal{A}$-coordinates. In fact the bipyramids themselves are the source of the non-prime initial ideals. The three non-prime initial ideals generated inside $b_1$ can be written as the intersection of two prime ideals as 
\begin{align*}
&\text{in}_{b_1 \setminus \{ b_{12} \}}(I_{3,6}) =  \langle \text{in}_{b_1 \setminus \{ b_{12} \}}(I_{3,6}) \cup M_1 \rangle \cap \langle \text{in}_{b_1 \setminus \{ b_{12} \}}(I_{3,6}) \cup \{ p_{256} p_{134} - p_{234} p_{156} \} \rangle,  \\
&\text{in}_{b_1 \setminus \{ b_{13} \}}(I_{3,6}) =  \langle \text{in}_{b_1 \setminus \{ b_{13} \}}(I_{3,6}) \cup M_1 \rangle \cap \langle \text{in}_{b_1 \setminus \{ b_{13} \}}(I_{3,6}) \cup \{ p_{124} p_{356} - p_{123} p_{456} \} \rangle,  \\
&\text{in}_{b_1 \setminus \{ b_{14} \}}(I_{3,6}) =  \langle \text{in}_{b_1 \setminus \{ b_{14} \}}(I_{3,6}) \cup M_1 \rangle \cap \langle \text{in}_{b_1 \setminus \{ b_{14} \}}(I_{3,6}) \cup \{ p_{126} p_{345}-p_{125} p_{346} \} \rangle.
\end{align*}


Similarly for $b_2$ we have the cyclic copy of the above given by 
\begin{align*}
&\text{in}_{b_2 \setminus \{ b_{22} \} }(I_{3,6}) =  \langle \text{in}_{b_2 \setminus \{ b_{22} \} }(I_{3,6}) \cup M_2 \rangle \cap \langle \text{in}_{b_2 \setminus \{ b_{22} \} }(I_{3,6}) \cup \{ p_{145} p_{236} - p_{123} p_{456} \} \rangle,  \\
&\text{in}_{b_2 \setminus \{ b_{23} \} }(I_{3,6}) =  \langle \text{in}_{b_2 \setminus \{ b_{23} \} }(I_{3,6}) \cup M_2 \rangle \cap \langle \text{in}_{b_2 \setminus \{ b_{23} \} }(I_{3,6}) \cup \{ p_{136} p_{245} - p_{126} p_{345} \} \rangle,  \\
&\text{in}_{b_2 \setminus \{ b_{24} \} }(I_{3,6}) =  \langle \text{in}_{b_2 \setminus \{ b_{24} \} }(I_{3,6}) \cup M_2 \rangle \cap \langle \text{in}_{b_2 \setminus \{ b_{24} \} }(I_{3,6}) \cup \{ p_{156} p_{234}-p_{146} p_{235} \} \rangle.
\end{align*}
In the above equations we understand $\text{in}_{b_1 \setminus \{ b_{11} \}}(I_{3,6})$ for instance, as the initial ideal of $I_{3,6}$ associated to the cone spanned by the rays $b_1 \setminus \{ b_{11} \}$ and we have defined the sets of monomials 
\begin{align*}
&M_1= \{p_{235},p_{236},p_{245},p_{246},p_{135},p_{136},p_{145},p_{146}\},\\
&M_2= \{p_{124},p_{125},p_{134},p_{135},p_{246},p_{256},p_{346},p_{356}\}.
\end{align*}
Most importantly notice the three quadratic non-prime factors appearing in each cone modulo the Pl\"ucker ideal are equivalent to either
\begin{align}
p_{ 12[34]56 } \text{ or } p_{ 23[45]61 },
\label{quadAs}
\end{align} 
where we have defined $p_{ij[kl]mn} = p_{ijl}p_{kmn}-p_{ijk}p_{lmn}$. These are exactly the two missing $\mathcal{A}$-coordinates, which along with the $14$ monomials contained in $M_1 \cup M_2$, make up the full set of active $\mathcal{A}$-coordinates.

The appearance of the quadratic $\mathcal{A}$-coordinates (\ref{quadAs}) suggests extending our original ideal by including them as new variables.
Extending the Pl\"ucker ideal as 
\be
I'_{3,6}=I_{3,6} \ \cap \langle q_1 - p_{ 12 [34] 56 } \rangle \subset \mathbb{R}[p_{123},\ldots,p_{456},q_1],
\ee
defines a new ideal and hence a new Gr\"obner fan $GF(I'_{3,6})$. The Speyer--Williams fan can also be refined by evaluating $q_1 = p_{ 12 [34] 56 }$\footnote{Similar to the notation in \eqref{Pluckerrels} $p_{ 12 [34] 56 }=p_{124}p_{356}-p_{123}p_{456}$.} in terms of minors of the web matrix and including it in the set of tropical polynomials defining the fan. In this way we obtain a refined Speyer--Williams fan with $f$-vector 
\be
f'_{3,6}=(1_0,16_1,66_2,99_3,48_4+1_5),
\ee
where we now have only a single five-vertex bipyramid ($b_2$). The bipyramid $b_1$ has been broken into two tetrahedra with an additional triangle separating them. Indeed, let us consider the natural extension of the map (\ref{evmap36})
\begin{align}
{\rm \bf ev}': (\tilde{x}_{11},\tilde{x}_{12},\tilde{x}_{21},\tilde{x}_{22}) \mapsto &\sum_{1\leq i < j < k \leq 6}  \tilde{p}_{ijk}(\tilde{x}_{11},\tilde{x}_{12},\tilde{x}_{21},\tilde{x}_{22}) e_{ijk} \notag \\
&+ \tilde{q}_1(\tilde{x}_{11},\tilde{x}_{12},\tilde{x}_{21},\tilde{x}_{22}) e_{q_{1}}
\label{evmap36'}
\end{align}
which takes a point in the refined Speyer--Williams fan into ${\rm Trop}^+(I'_{3,6})$, where we introduced $e_{q_{1}}$ as a unit vector for the $q_1$ direction. Under this map the images of the five rays of $b_1$ are now \cite{Drummond:2020kqg}
\begin{align}
b_{11} &\mapsto e_{123} + e_{124} + e_{125} + e_{126} + e_{134} + e_{234} + e_{q_1}\,, \notag \\
b_{12} &\mapsto  e_{123} + e_{124} + e_{134} + e_{234} + e_{q_1} \,, \notag \\
b_{13} &\mapsto e_{125} + e_{126} + e_{156} + e_{256} + e_{q_1} \,, \notag \\
b_{14} &\mapsto e_{345} + e_{346} + e_{356} + e_{456} + e_{q_1} \,, \notag \\
b_{15} &\mapsto e_{123} + e_{124} + e_{134} + e_{234} + e_{345} + e_{346} + e_{q_1}\,,
\label{eq1ext}
\end{align}
and the relation (\ref{bipyramidcentre}) no longer holds if we replace ${\rm \bf ev}$ with ${\rm \bf ev}'$.

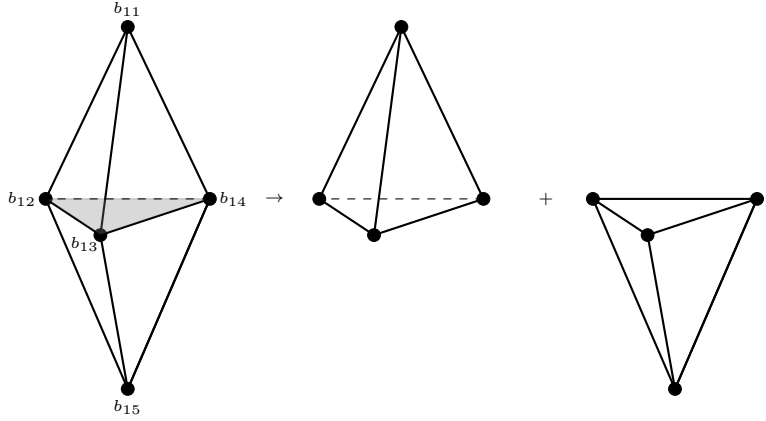
\begin{figure}[h]

\centering

\begin{tikzpicture}[scale=1.2]

   \draw[thick] (3.25,0.1) node[right] {{\tiny $ \rightarrow $}};

    \draw[fill=black] (1.8,2) circle (0.07);

     \draw[fill=black] (1.8,-2) circle (0.07);

\draw[fill=black] ($(1.8,0)+(-0.9,0.1)$) node[left] {{\tiny $b_{12}$}\,}  circle (0.07);

       \draw[fill=black] ($(1.8,0)+(+0.9,0.1)$) node[right] {\,{\tiny $b_{14}$}}  circle (0.07);

        \draw[fill=black] ($(1.8,0)+(-0.3,-0.3)$) circle (0.07);

    \draw[thick] (1.8,2) -- ($(1.8,0)+(0.9,0.1)$);
\node[] at (1.8,2.2) {{\tiny $b_{11}$}} ;

    \draw[thick] (1.8,-2)  -- ($(1.8,0)+(0.9,0.1)$); 
\node[] at (1.8,-2.2) {{\tiny $b_{15}$}};

    \draw[thick] (1.8,2) -- ($(1.8,0)+(-0.9,0.1)$);

    \draw[thick] (1.8,2) -- ($(1.8,0)+(-0.3,-0.3)$);

    \draw[thick] ($(1.8,0)+(-0.3,-0.3)$) --($(1.8,0)+(0.9,0.1)$);

    \draw[thick] (1.8,-2) -- ($(1.8,0)+(-0.3,-0.3)$);

    \draw[thick] ($(1.8,0)+(-0.9,0.1)$) -- ($(1.8,0)+(-0.3,-0.3)$);

    \draw[line width=0, fill=gray, opacity=0.3] ($(1.8,0)+(-0.9,0.1)$) -- ($(1.8,0)+(+0.9,0.1)$) -- ($(1.8,0)+(-0.3,-0.3)$) -- cycle;

    \draw[dashed] ($(1.8,0)+(0.9,0.1)$) -- ($(1.8,0)+(-0.9,0.1)$);

    \draw[thick] (1.8,-2)  -- ($(1.8,0)+(0.9,0.1)$);

    \draw[thick] (1.8,-2) -- ($(1.8,0)+(-0.9,0.1)$);

    \node[] at (1.33,-0.4) {\tiny $ b_{13}$};


    \draw[thick] (6.25,0.1) node[right] {{\tiny $ + $}};

   \draw[fill=black] (4.8,2) circle (0.07);

   \draw[fill=black] ($(4.8,0)+(-0.9,0.1)$)  circle (0.07);

       \draw[fill=black] ($(4.8,0)+(+0.9,0.1)$)  circle (0.07);

        \draw[fill=black] ($(4.8,0)+(-0.3,-0.3)$) circle (0.07);

    \draw[thick] (4.8,2) -- ($(4.8,0)+(0.9,0.1)$);

    \draw[thick] (4.8,2) -- ($(4.8,0)+(-0.9,0.1)$);

    \draw[thick] (4.8,2) -- ($(4.8,0)+(-0.3,-0.3)$);

    \draw[thick] ($(4.8,0)+(-0.3,-0.3)$) --($(4.8,0)+(0.9,0.1)$);

    \draw[thick] ($(4.8,0)+(-0.9,0.1)$) -- ($(4.8,0)+(-0.3,-0.3)$);

    \draw[dashed] ($(4.8,0)+(0.9,0.1)$) -- ($(4.8,0)+(-0.9,0.1)$);


     \draw[fill=black] (7.8,-2) circle (0.07);

   \draw[fill=black] ($(7.8,0)+(-0.9,0.1)$)  circle (0.07);

       \draw[fill=black] ($(7.8,0)+(+0.9,0.1)$)  circle (0.07);

        \draw[fill=black] ($(7.8,0)+(-0.3,-0.3)$) circle (0.07);

    \draw[thick] (7.8,-2)  -- ($(7.8,0)+(0.9,0.1)$);

    \draw[thick] ($(7.8,0)+(-0.3,-0.3)$) --($(7.8,0)+(0.9,0.1)$);

    \draw[thick] (7.8,-2) -- ($(7.8,0)+(-0.3,-0.3)$);

    \draw[thick] ($(7.8,0)+(-0.9,0.1)$) -- ($(7.8,0)+(-0.3,-0.3)$);

    \draw[thick] ($(7.8,0)+(0.9,0.1)$) -- ($(7.8,0)+(-0.9,0.1)$);

    \draw[thick] (7.8,-2)  -- ($(7.8,0)+(0.9,0.1)$);

    \draw[thick] (7.8,-2) -- ($(7.8,0)+(-0.9,0.1)$);

\end{tikzpicture}

\caption{\small The bipyramid $b_1$ maps to two tetrahedra inside $\text{Trop}^+(I'_{3,6})$, both of which are associated to prime ideals. }
\label{fig:grob_split_36_2}

\end{figure}


The transition from $f_{3,6}$ to $f'_{3,6}$ can be seen as adding a triangle to the equator of the bipyramid $b_1$ as shown in  Figure \ref{fig:grob_split_36_2} which now splits into two tetrahedra whose images in ${\rm Trop}^+(I'_{3,6})$ now correspond to prime initial ideals. In other words, at least as far as these five rays are concerned, the fan structures of the extended Speyer--Williams fan and the positive tropical fan of $I'_{3,6}$ match. Also we find that the sixteen rays of $\text{Trop}^+(I'_{3,6})$ now span only $3$ maximal Gr\"obner cones inside $GF(I'_{3,6})$.

Extending the ideal further as 
\be
I^{\text{ext}}_{3,6}=I'_{3,6} \  \cap \langle q_2-p_{ 23 [45] 61 } \rangle \subset \mathbb{R}[p_{123},\ldots,p_{456},q_1,q_2],
\ee
defines yet again a new Gr\"obner fan $GF(I^{\text{ext}}_{3,6})$. We can further refine the Speyer--Williams fan by tropically evaluating $q_2 = p_{ 23 [45] 61 }$. This fan was referred to as the cluster fan in \cite{Drummond:2020kqg} because it coincides with the g-vector fan of the ${\rm Gr}(3,6)$ cluster algebra. The cluster fan has f-vector 
\be
f^{\text{ext}}_{3,6}=(1_0,16_1,66_2,100_3,50_4)
\ee
and now its fan structure coincides with that of $\text{Trop}^+(I^{\text{ext}}_{3,6})$ which is the image of the cluster fan under the natural extension of (\ref{evmap36'}),
\begin{align}
\label{evmap36ext}
{\rm \bf ev}^{\rm ext}: (\tilde{x}_{11},\tilde{x}_{12},\tilde{x}_{21},\tilde{x}_{22}) \mapsto &\sum_{1\leq i < j < k \leq 6}  \tilde{p}_{ijk}(\tilde{x}_{11},\tilde{x}_{12},\tilde{x}_{21},\tilde{x}_{22}) e_{ijk}  \\
&+ \tilde{q}_1(\tilde{x}_{11},\tilde{x}_{12},\tilde{x}_{21},\tilde{x}_{22}) e_{q_{1}} + \tilde{q}_2(\tilde{x}_{11},\tilde{x}_{12},\tilde{x}_{21},\tilde{x}_{22}) e_{q_{2}}\,. \notag
\end{align}
Note that $\text{Trop}^+(I^{\text{ext}}_{3,6})$ contains no non-prime maximal cones! The effect of the further extension can be viewed as adding a triangle to the equator of bipyramid $b_2$ in an exact copy of Figure \ref{fig:grob_split_36_2}, while the rays of $b_2$ are extended in the new $e_{q_2}$ direction in an analogous manner to (\ref{eq1ext}).

Furthermore, the rays of $\text{Trop}^+(I^{\text{ext}}_{3,6})$ now span a single maximal Gr\"obner cone whose initial ideal is generated by
\begin{align*}
\{&q_1 q_2,p_{124} q_2,p_{125} q_2,p_{134} q_2,p_{135} q_1,p_{135} q_2,p_{124} p_{135},p_{136} q_1,p_{124} p_{136},p_{125} p_{136},p_{145} q_1, \\
&p_{146} q_1,p_{125} p_{146},p_{135} p_{146},p_{235} q_1,p_{124} p_{235},p_{134} p_{235},p_{146} p_{235},p_{236} q_1,p_{124} p_{236},p_{125} p_{236}, \\
&p_{134} p_{236},p_{135} p_{236},p_{145} p_{236},p_{245} q_1,p_{134} p_{245},p_{135} p_{245},p_{136} p_{245},p_{246} q_1,p_{246} q_2,p_{125} p_{246},\\
&p_{134} p_{246},p_{135} p_{246},p_{136} p_{246},p_{145} p_{246},p_{235} p_{246},p_{256} q_2,p_{134} p_{256},p_{135} p_{256},p_{136} p_{256},p_{145} p_{256},\\
&p_{146} p_{256},p_{346} q_2,p_{125} p_{346},p_{135} p_{346},p_{145} p_{346},p_{235} p_{346},p_{245} p_{346},p_{356} q_2,p_{124} p_{356},p_{145} p_{356},\\
&p_{146} p_{356},p_{245} p_{356},p_{246} p_{356}\},
\end{align*}
which are exactly the $54$ forbidden pairs of $\mathcal{A}$-coordinates (i.e. those pairs which never sit together in a cluster). 

In summary, we have seen explicitly that the non-prime initial ideals provided the missing quadratic $\mathcal{A}$ coordinates. These new variables also tell us how to extend the Pl\"ucker ideal $I_{3,6}$ to a new ideal $I_{3,6}^{\rm ext}$ so as to find a single maximal Gr\"obner cone inside the span of the rays of the corresponding positive tropical fan ${\rm Trop}^+(I_{3,6}^{\rm ext})$. The initial ideal associated to this maximal Gr\"obner cone provides the adjacency information of the cluster algebra. Thus these two key pieces of information ($\mathcal{A}$-coordinates and adjacency) can be obtained from the Gr\"obner structure of the relevant ideals without direct reference to the cluster algebra at all.

\subsection{${\rm Gr}(3,7) \cong {\rm Gr}(4,7)$}
 We now go beyond the results of \cite{boss2021grob} and consider ${\rm Gr}(3,7) \cong {\rm Gr}(4,7)$. We again wish to compare the structure of the positive tropical fan $\text{Trop}^+(I_{3,7})$ to the Speyer--Williams fan obtained from tropicalising the $35$ Pl\"ucker coordinates evaluated as minors of the web matrix.

Let us begin by detailing the structure of the full tropical fan $\text{Trop}(I_{3,7})$ which was calculated in \cite{HJJS,Jensenwebpage}. The  tropical fan $\text{Trop}(I_{3,7})$ is simplicial with an $f$-vector given by
\be
(1,721,16800,124180,386155,522585,252000)\,.
\ee
where the $721$ rays are given by the permutation copies of \cite{Cachazo:2019apa}
  \begin{align}
      b_{1,1234567} & = e_{123}, \notag \\
    b_{2,1234567} & = e_{123}+e_{124}+e_{134}+e_{234}, \notag\\
    b_{3,1234567} & = e_{123}+e_{124}+e_{125}+e_{126}+e_{127}, \notag \\
    b_{4,1234567} & = e_{123}+e_{124}+e_{125}+e_{126}+e_{127}+e_{134}+e_{234}, \notag \\
    b_{5,1234567} & = e_{123}+e_{124}+e_{125}+e_{126}+e_{127}+e_{134}+e_{156}+e_{234}+e_{256}, \notag \\
    b_{6,1234567} & = b_{3,1234567}+b_{3,3456712}+b_{3_,6712345}.
    \label{eq:bee6}
\end{align}
The positive part of the tropical fan $\text{Trop}^+(I_{3,7})$ is spanned by $49$ rays given by the cyclic copies of 
\be
\{ b_{1,1234567},b_{2,1234567},b_{3,1234567} ,b_{4,1234567} ,b_{4,1562347} ,b_{5,1234675} ,b_{6,1234567}  \}.
\label{eq:pos_vecs}
\ee
By restricting to the subfan spanned by these rays we obtain the positive tropical fan $\text{Trop}^+(I_{3,7})$ with the $f$-vector
\be
(1,49,490,1964,3633,3192,1064).
\ee

Now we turn our attention to the Speyer--Williams fan \cite{speyer2003tropical} whose structure is summarised by the f-vector 
\be
f_{3,7}=(1,42,392,1463,2583,2163,595_6 + 63_7 + 28_8 + 7_9),
\ee
where we have included the information on the number of vertices for the maximal cones only (in the final element above). This fan is spanned by $42$ rays given by the first six elements of \eqref{eq:pos_vecs} along with their cyclic copies and has $6$-dimensional maximal cones. Note that ${\rm Trop}^+(I_{3,7})$ has seven more rays than the Speyer--Williams fan. These additional rays correspond to $b_{6,1234567}$ and cyclic copies. As discussed in \cite{Cachazo:2019apa,Drummond:2019qjk} these extra rays actually appear in the middle of triangular faces formed by three rays of $b_3$-type.

As discussed in previous cases, we can map the Speyer--Williams fan to the positive tropical Grassmannian ${\rm Trop}^+(I_{3,7})$ via the evaluation map,
\be
{\rm \bf ev}: (\tilde{x}_{11},\ldots,\tilde{x}_{23}) \mapsto \sum \tilde{p}_{ijk}(\tilde{x}_{11},\ldots,\tilde{x}_{23}) e_{ijk}\,.
\ee
Once again we find that the fan structure of ${\rm Trop}^+(I_{3,7})$ is a refinement of the (image of the) Speyer--Williams fan. Let us explore this refinement in more detail.

\begin{figure}[h]
\centering
\begin{tikzpicture}[scale=1.7]
    
   \draw[thick] (3.25,0.1) node[right] {{\tiny $ = $}};
    
    \draw[fill=DarkGreen] (1.8,2) circle (0.04);

    \draw[thick] (1.8,2) node[above] {{\tiny $ $}} -- ($(1.8,0)+(0.9,0.1)$);
    \draw[thick] (1.8,2) -- ($(1.8,0)+(-0.9,0.1)$);
    
    \draw[thick] (1.8,2) -- ($(1.8,0)+(-0.3,-0.3)$);
    \draw[thick] ($(1.8,0)+(-0.3,-0.3)$) --($(1.8,0)+(0.9,0.1)$);
    \draw[thick] ($(1.8,0)+(-0.9,0.1)$) -- ($(1.8,0)+(-0.3,-0.3)$);

    \draw[dashed] ($(1.8,0)+(0.9,0.1)$) -- ($(1.8,0)+(-0.9,0.1)$);

     \draw[dashed] (1.8,2) node[above] {{\tiny $ $}} -- ($(1.7, -0.0333333)$);
    \draw[dashed] (2.7,0.1) node[above] {{\tiny $ $}} -- ($(1.7, -0.0333333)$);
    \draw[dashed] ($(1.8,0)+(-0.9,0.1)$) node[above] {{\tiny $ $}} -- ($(1.7, -0.0333333)$);
    \draw[dashed] (1.5,-0.3) node[above] {{\tiny $ $}} -- ($(1.7, -0.0333333)$);
        
        \node[] at (1.33,-0.5) {\tiny $  $};
    \draw[fill=DarkGreen] (1.8,2) circle (0.05);
     \draw[fill=red] ($(1.8,0)+(-0.9,0.1)$) node[left] {{\tiny $ $}}  circle (0.05);
       \draw[fill=blue] ($(1.8,0)+(+0.9,0.1)$) node[right] {{\tiny $ $}}  circle (0.05);
        \draw[fill=pink] ($(1.8,0)+(-0.3,-0.3)$) circle (0.05);
        
         \draw[fill=orange] ($(1.7, -0.0333333)$) node[left] {{\tiny $ $}}  circle (0.05);

    
     \draw[thick] (5.25,0.1) node[right] {{\tiny $ + $}};

       \draw[thick] (4.8,2) node[above] {{\tiny $ $}} -- ($(4.7, -0.0333333)$);
    \draw[dashed] (3.9,0.1) node[above] {{\tiny $ $}} -- ($(4.7, -0.0333333)$);
    \draw[thick] (4.5,-0.3) node[above] {{\tiny $ $}} -- ($(4.7, -0.0333333)$);
    
    \draw[thick] (4.8,2) -- ($(4.8,0)+(-0.9,0.1)$);
    
    \draw[thick] (4.8,2) -- ($(4.8,0)+(-0.3,-0.3)$);
    \draw[thick] ($(4.8,0)+(-0.9,0.1)$) -- ($(4.8,0)+(-0.3,-0.3)$);

    
    \node[] at (4.33,-0.4) {\tiny $ $};
    \draw[fill=DarkGreen] (4.8,2) circle (0.05);
     \draw[fill=red] ($(4.8,0)+(-0.9,0.1)$) node[left] {{\tiny $ $}}  circle (0.05);
        \draw[fill=pink] ($(4.8,0)+(-0.3,-0.3)$) circle (0.05);
           \draw[fill=orange] ($(4.7, -0.0333333)$) node[left] {{\tiny $ $}}  circle (0.05);
           
           \draw[fill=orange] ($(4.7, -0.5333333)$) node {{\tiny $ p_{124} p_{367} - p_{123} p_{467} $}} ;
           

     \draw[thick] (8.1,0.1) node {{\tiny $ + $}};

    \draw[thick] (6.8,2) node[above] {{\tiny $ $}} -- ($(6.7, -0.0333333)$);
    \draw[thick] (7.7,0.1) node[above] {{\tiny $ $}} -- ($(6.7, -0.0333333)$);
    \draw[thick] (5.9,0.1) node[above] {{\tiny $ $}} -- ($(6.7, -0.0333333)$);
    
    \draw[thick] (6.8,2) node[above] {{\tiny $ $}} -- ($(6.8,0)+(0.9,0.1)$);
    \draw[thick] (6.8,2) -- ($(6.8,0)+(-0.9,0.1)$);
    

    \draw[dashed] ($(6.8,0)+(0.9,0.1)$) -- ($(6.8,0)+(-0.9,0.1)$);
    
    \node[] at (6.33,-0.4) {\tiny $  $};
    \draw[fill=DarkGreen] (6.8,2) circle (0.05);
     \draw[fill=red] ($(6.8,0)+(-0.9,0.1)$) node[left] {{\tiny $ $}}  circle (0.05);
       \draw[fill=blue] ($(6.8,0)+(+0.9,0.1)$) node[right] {{\tiny $ $}}  circle (0.05);
         \draw[fill=orange] ($(6.7, -0.0333333)$) node[left] {{\tiny $ $}}  circle (0.05);
    
	\draw[fill=orange] ($(6.7, -0.5333333)$) node {{\tiny $ p_{167} p_{234} - p_{134} p_{267}$}} ;


    \draw[dashed] (8.8,2) node[above] {{\tiny $ $}} -- ($(8.7, -0.0333333)$);
    \draw[dashed] (9.7,0.1) node[above] {{\tiny $ $}} -- ($(8.7, -0.0333333)$);
    \draw[dashed] (8.5,-0.3) node[above] {{\tiny $ $}} -- ($(8.7, -0.0333333)$);
    
    \draw[thick] (8.8,2) node[above] {{\tiny $ $}} -- ($(8.8,0)+(0.9,0.1)$);
    
    \draw[thick] (8.8,2) -- ($(8.8,0)+(-0.3,-0.3)$);
    \draw[thick] ($(8.8,0)+(-0.3,-0.3)$) --($(8.8,0)+(0.9,0.1)$);

    
    \node[] at (8.33,-0.4) {\tiny $ $};
    \draw[fill=DarkGreen] (8.8,2) circle (0.05);
       \draw[fill=blue] ($(8.8,0)+(+0.9,0.1)$) node[right] {{\tiny $ $}}  circle (0.05);
        \draw[fill=pink] ($(8.8,0)+(-0.3,-0.3)$) circle (0.05);
               \draw[fill=orange] ($(8.7, -0.0333333)$) node[left] {{\tiny $ $}}  circle (0.05);
               
               \draw[fill=orange] ($(8.7, -0.5333333)$) node {{\tiny $ p_{127} p_{346} - p_{126} p_{347}$}} ;
           
 \end{tikzpicture}
\caption{\small On the left a $6$-vertex simplex of the Speyer--Williams fan. On the right the refinement of the Speyer--Williams cone by the positive tropical fan $\text{Trop}^+(I_{3,7})$. Note the appearance of a new orange vertex.}
\label{fig:6vertexsplit}
\end{figure}
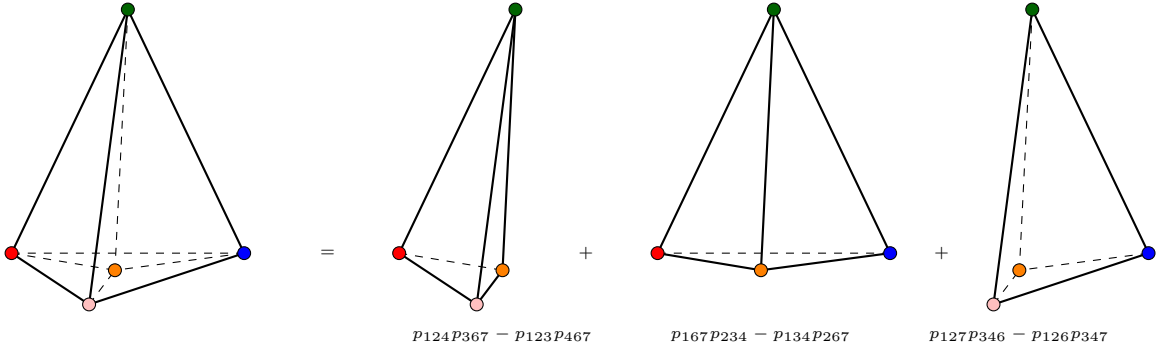

Let us look at each of the maximal cones in turn starting with the six-vertex cones of the Speyer--Williams fan. 
There $595$ simplicial maximal cones of the Speyer--Williams fan: $567$ are un-refined by the positive tropical fan, and $28$ are refined by the positive tropical fan coming in $2$ dihedral classes. As an example, a representative from the first dihedral class has the rays given by
\be
\left\{\textcolor{DarkGreen}{b_{\text{1,5671234}}},\textcolor{DarkGreen}{b_{\text{4,1234567}}},\textcolor{DarkGreen}{b_{\text{4,6712345}}},\textcolor{red}{b_{\text{3,6712345}}},\textcolor{blue}{b_{\text{3,3456712}}},\textcolor{pink}{b_{\text{3,1234567}}}\right\}
\ee
and is refined by the tropical fan with three simplices given by 
\be
\left(
\begin{array}{cccccc}
 \textcolor{DarkGreen}{b_{\text{1,5671234}}} & \textcolor{DarkGreen}{b_{\text{4,1234567}}} &  \textcolor{DarkGreen}{b_{\text{4,6712345}}} & \textcolor{red}{b_{\text{3,6712345}}} & \textcolor{pink}{b_{\text{3,1234567}}}  & \textcolor{orange}{b_{\text{6,1234567}}} \\
 \textcolor{DarkGreen}{b_{\text{1,5671234}}} & \textcolor{DarkGreen}{b_{\text{4,1234567}}} &  \textcolor{DarkGreen}{b_{\text{4,6712345}}}  & \textcolor{red}{b_{\text{3,6712345}}} & \textcolor{blue}{b_{\text{3,3456712}}}  & \textcolor{orange}{b_{\text{6,1234567}}} \\
 \textcolor{DarkGreen}{b_{\text{1,5671234}}} & \textcolor{DarkGreen}{b_{\text{4,1234567}}} &  \textcolor{DarkGreen}{b_{\text{4,6712345}}} & \textcolor{blue}{b_{\text{3,3456712}}} & \textcolor{pink}{b_{\text{3,1234567}}}  & \textcolor{orange}{b_{\text{6,1234567}}} \\
\end{array}
\right).
\ee
Note, in particular the appearance of the extra positive ray $b_{\text{6,1234567}}$, which is a ray of ${\rm Trop}^+(I_{3,7})$ but not of the Speyer--Williams fan. We give a sketch showing the appearance of the extra ray in the middle of a face as shown in Figure \ref{fig:6vertexsplit}. Below each part of the refinement in Figure \ref{fig:6vertexsplit} we note the missing binomial contained as a factor in the non-prime initial ideal associated to this cone of ${\rm Trop}^+(I_{3,7})$.

\begin{figure}[h]
\centering
\begin{tikzpicture}[scale=1.7]
    
   \draw[thick] (3.25,0.1) node[right] {{\tiny $ = $}};
    
    \draw[fill=DarkGreen] (1.8,2) circle (0.04);

    \draw[thick] (1.8,2) node[above] {{\tiny $ $}} -- ($(1.8,0)+(0.9,0.1)$);
    \draw[thick] (1.8,2) -- ($(1.8,0)+(-0.9,0.1)$);
    
    \draw[thick] (1.8,2) -- ($(1.8,0)+(-0.3,-0.3)$);
    \draw[thick] ($(1.8,0)+(-0.3,-0.3)$) --($(1.8,0)+(0.9,0.1)$);
    \draw[thick] ($(1.8,0)+(-0.9,0.1)$) -- ($(1.8,0)+(-0.3,-0.3)$);

    \draw[dashed] ($(1.8,0)+(0.9,0.1)$) -- ($(1.8,0)+(-0.9,0.1)$);

     \draw[dashed] (1.8,2) node[above] {{\tiny $ $}} -- ($(1.7, -0.0333333)$);
    \draw[dashed] (2.7,0.1) node[above] {{\tiny $ $}} -- ($(1.7, -0.0333333)$);
    \draw[dashed] ($(1.8,0)+(-0.9,0.1)$) node[above] {{\tiny $ $}} -- ($(1.7, -0.0333333)$);
    \draw[dashed] (1.5,-0.3) node[above] {{\tiny $ $}} -- ($(1.7, -0.0333333)$);
        
         \node[] at (1.33,-0.5) {\tiny $  $};
    \draw[fill=DarkGreen] (1.8,2) circle (0.05);
     \draw[fill=red] ($(1.8,0)+(-0.9,0.1)$) node[left] {{\tiny $ $}}  circle (0.05);
       \draw[fill=blue] ($(1.8,0)+(+0.9,0.1)$) node[right] {{\tiny $ $}}  circle (0.05);
        \draw[fill=pink] ($(1.8,0)+(-0.3,-0.3)$) circle (0.05);
        
         \draw[fill=orange] ($(1.7, -0.0333333)$) node[left] {{\tiny $ $}}  circle (0.05);

    
     \draw[thick] (5.25,0.1) node[right] {{\tiny $ + $}};

       \draw[thick] (4.8,2) node[above] {{\tiny $ $}} -- ($(4.7, -0.0333333)$);
    \draw[dashed] (3.9,0.1) node[above] {{\tiny $ $}} -- ($(4.7, -0.0333333)$);
    \draw[thick] (4.5,-0.3) node[above] {{\tiny $ $}} -- ($(4.7, -0.0333333)$);
    
    \draw[thick] (4.8,2) -- ($(4.8,0)+(-0.9,0.1)$);
    
    \draw[thick] (4.8,2) -- ($(4.8,0)+(-0.3,-0.3)$);
    \draw[thick] ($(4.8,0)+(-0.9,0.1)$) -- ($(4.8,0)+(-0.3,-0.3)$);

    
    \node[] at (4.33,-0.4) {\tiny $ $};
    \draw[fill=DarkGreen] (4.8,2) circle (0.05);
     \draw[fill=red] ($(4.8,0)+(-0.9,0.1)$) node[left] {{\tiny $ $}}  circle (0.05);
        \draw[fill=pink] ($(4.8,0)+(-0.3,-0.3)$) circle (0.05);
           \draw[fill=orange] ($(4.7, -0.0333333)$) node[left] {{\tiny $ $}}  circle (0.05);
           
           \draw[fill=orange] ($(4.7, -0.5333333)$) node {{\tiny $ p_{124} p_{367} - p_{123} p_{467} $}} ;
           

     \draw[thick] (8.1,0.1) node {{\tiny $ + $}};

    \draw[thick] (6.8,2) node[above] {{\tiny $ $}} -- ($(6.7, -0.0333333)$);
    \draw[thick] (7.7,0.1) node[above] {{\tiny $ $}} -- ($(6.7, -0.0333333)$);
    \draw[thick] (5.9,0.1) node[above] {{\tiny $ $}} -- ($(6.7, -0.0333333)$);
    
    \draw[thick] (6.8,2) node[above] {{\tiny $ $}} -- ($(6.8,0)+(0.9,0.1)$);
    \draw[thick] (6.8,2) -- ($(6.8,0)+(-0.9,0.1)$);
    

    \draw[dashed] ($(6.8,0)+(0.9,0.1)$) -- ($(6.8,0)+(-0.9,0.1)$);
    
    \node[] at (6.33,-0.4) {\tiny $  $};
    \draw[fill=DarkGreen] (6.8,2) circle (0.05);
     \draw[fill=red] ($(6.8,0)+(-0.9,0.1)$) node[left] {{\tiny $ $}}  circle (0.05);
       \draw[fill=blue] ($(6.8,0)+(+0.9,0.1)$) node[right] {{\tiny $ $}}  circle (0.05);
         \draw[fill=orange] ($(6.7, -0.0333333)$) node[left] {{\tiny $ $}}  circle (0.05);
    
	\draw[fill=orange] ($(6.7, -0.5333333)$) node {{\tiny $ p_{167} p_{234} - p_{134} p_{267}$}} ;


    \draw[dashed] (8.8,2) node[above] {{\tiny $ $}} -- ($(8.7, -0.0333333)$);
    \draw[dashed] (9.7,0.1) node[above] {{\tiny $ $}} -- ($(8.7, -0.0333333)$);
    \draw[dashed] (8.5,-0.3) node[above] {{\tiny $ $}} -- ($(8.7, -0.0333333)$);
    
    \draw[thick] (8.8,2) node[above] {{\tiny $ $}} -- ($(8.8,0)+(0.9,0.1)$);
    
    \draw[thick] (8.8,2) -- ($(8.8,0)+(-0.3,-0.3)$);
    \draw[thick] ($(8.8,0)+(-0.3,-0.3)$) --($(8.8,0)+(0.9,0.1)$);

    
    \node[] at (8.33,-0.4) {\tiny $ $};
    \draw[fill=DarkGreen] (8.8,2) circle (0.05);
       \draw[fill=blue] ($(8.8,0)+(+0.9,0.1)$) node[right] {{\tiny $ $}}  circle (0.05);
        \draw[fill=pink] ($(8.8,0)+(-0.3,-0.3)$) circle (0.05);
               \draw[fill=orange] ($(8.7, -0.0333333)$) node[left] {{\tiny $ $}}  circle (0.05);
               
               \draw[fill=orange] ($(8.7, -0.5333333)$) node {{\tiny $ p_{127} p_{346} - p_{126} p_{347}$}} ;
           
 \end{tikzpicture}
\caption{\small The second dihedral class of cones split by an extra positive ray of $b_6$ type.}
\label{fig:6vertexsplit2}
\end{figure}
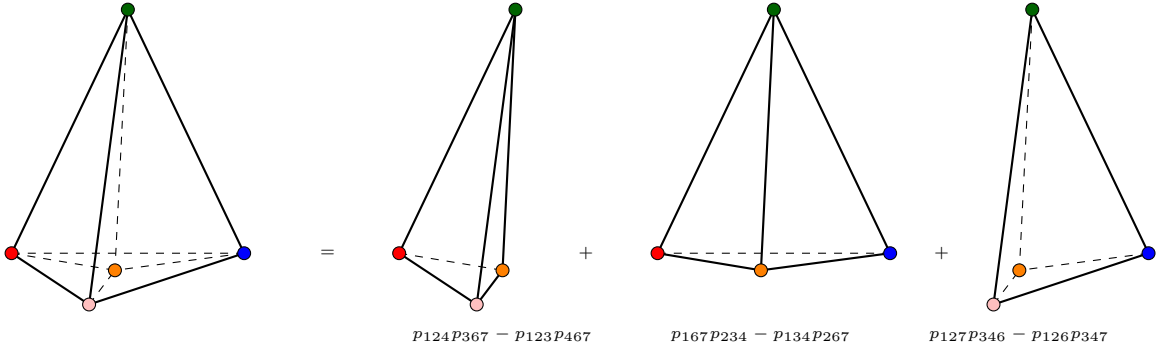

The second dihedral set of $6$-vertex cones which are refined by positive tropical fan have the rays (in the Speyer--Williams fan)  given by
\be
\left\{\textcolor{DarkGreen}{b_{\text{1,5671234}}},\textcolor{DarkGreen}{b_{\text{1,3456712}}},\textcolor{DarkGreen}{b_{\text{4,6712345}}},\textcolor{red}{b_{\text{3,6712345}}},\textcolor{blue}{b_{\text{3,3456712}}}, \textcolor{pink}{b_{\text{3,1234567}}}\right\}
\ee 
which are split into three simplices (in the positive tropical fan) with the rays
\be
\left(
\begin{array}{cccccc}
 \textcolor{DarkGreen}{b_{\text{1,5671234}}}& \textcolor{DarkGreen}{b_{\text{1,3456712}}} & \textcolor{DarkGreen}{b_{\text{4,6712345}}} & \textcolor{red}{b_{\text{3,6712345}}} & \textcolor{pink}{b_{\text{3,1234567}}} & \textcolor{orange}{b_{\text{6,1234567}}} \\
\textcolor{DarkGreen}{b_{\text{1,5671234}}} & \textcolor{DarkGreen}{b_{\text{1,3456712}}} & \textcolor{DarkGreen}{b_{\text{4,6712345}}} & \textcolor{red}{b_{\text{3,6712345}}} & \textcolor{blue}{b_{\text{3,3456712}}} & \textcolor{orange}{b_{\text{6,1234567}}} \\
 \textcolor{DarkGreen}{b_{\text{1,5671234}}} & \textcolor{DarkGreen}{b_{\text{1,3456712}}} & \textcolor{DarkGreen}{b_{\text{4,6712345}}} & \textcolor{blue}{b_{\text{3,3456712}}} & \textcolor{pink}{b_{\text{3,1234567}}} & \textcolor{orange}{b_{\text{6,1234567}}} \\
\end{array}
\right)\,.
\ee
Again we note the appearance of a spurious positive ray. The picture for this cone looks similar and is illustrated in Figure \ref{fig:6vertexsplit2}, with the same missing binomials appearing.

The remaining $6$-vertex cones coincide in both fans, i.e. are unrefined. The remaining maximal cones have 7, 8 or 9 vertices. We explore their refinements in Appendix \ref{app:Gr37}, with particular focus on the `missing binomials' appearing as factors of non-prime initial ideals.

In summary, we find all $14$ missing quadratic $\mathcal{A}$-coordinates (with each one arising as a factor of non-prime initial ideals in many cones).
\begin{align}
q_{51}=p_{12[34]56} \text{ and cyclic},\notag\\ 
q_{61}=p_{61[23]45} \text{ and cyclic},
\end{align}
appearing as non-prime factors of the $7_9$ maximal cones. By extending the ideal by all $14$ variables as 
\be
I_{3,7}^{\text{ext}} = I_{3,7} \cap \langle  q_{51}-  p_{12[34]56}, \ldots,  q_{67}- p_{56[12]34}  \rangle \subset \mathbb{R}[p_{123},\ldots,p_{567},q_{51},\ldots, q_{67}] ,
\ee
the extended Speyer--Williams fan (again the cluster fan) and the positive tropical fan $\text{Trop}^+(I_{3,7}^{\text{ext}})$ become identical simplicial fans with the $f$-vector
$$f^{\text{ext}}_{3,7}=(1,42, 399, 1547, 2856, 2499, 833).$$
Moreover, the rays of $\text{Trop}^+(I_{3,7}^{\text{ext}})$ span a single maximal Gr\"obner cone in $GF(I_{3,7}^{\rm ext})$ whose initial ideal gives us precisely the list of \underline{$462$ forbidden neighbours} for $7$-point scattering amplitudes as given in \cite{Drummond:2017ssj}.

\subsection{${\rm Gr}(3,8)$}
In the space of variables $(\tilde{x}_{11},\tilde{x}_{12},\tilde{x}_{13},\tilde{x}_{14},\tilde{x}_{21},\tilde{x}_{22},\tilde{x}_{23},\tilde{x}_{24})$ the Speyer-Willliams fan is spanned by $120$ rays with $8$-dimensional maximal cones. The  structure of the fan is summarised by the f-vector
$$f_{3,8}=(1, 120, 2072, 14088, 48544, 93104, 100852, 57768, 13612),$$
with the 13612 maximal cones given by
$$
9672_8 + 1696_9 + 1092_{10} + 480_{11} + 416_{12} + 104_{13} + 
 88_{14} + 32_{15} + 24_{16} + 8_{17}.
$$
For the case of ${\rm Gr}(3,8)$ there two new features which we wish to emphasise. First, there are 8 more $\mathcal{A}$-coordinates than there are rays of the Speyer--Williams fan, which must appear when we begin to extend the ideal. Second, the $\mathcal{A}$-coordinates not only contain expressions quadratic in the Pl\"uckers but also contain cubic expressions given by
\be
\{ p_{12[34]5[67]89},p_{12[35]8[67]45},p_{12[34]8[67]35} \} \text{ and cylic},
\ee
where we have made the definition $p_{ij[kl]m[nr]st} = p_{ijl}p_{km[nr]st} - p_{ijk}p_{lm[nr]st}$.

Up to this point we have been searching the entire positive part of the tropical fan for the appearance of non-prime factors. However, for ${\rm Gr}(3,8)$ this calculation becomes cumbersome and instead we satisfy ourselves with a local approach. 
Concretely, we begin by searching for non-prime factors appearing in the $17$-vertex cone spanned by the rays 
\begin{equation}
\left(
\begin{smallmatrix}
 \textcolor{blue}{1} & \textcolor{blue}{2} & \textcolor{blue}{3} & \textcolor{blue}{4} & \textcolor{blue}{5} & \textcolor{blue}{6} & \textcolor{blue}{7} & \textcolor{blue}{8} & \textcolor{blue}{9} & \textcolor{blue}{10} & \textcolor{blue}{11} & \textcolor{blue}{12} & \textcolor{blue}{13} & \textcolor{blue}{14} & \textcolor{blue}{15} & \textcolor{blue}{16} & \textcolor{blue}{17} & \\
   & &  &  &  &  &  &  &  &  & &  &  &  &  &  &  \\
  \hline\\
    & &  &  &  &  &  &  &  &  & &  &  &  &  &  &  \\
 1 & 1 & 1 & 1 & 1 & 1 & 1 & 1 & 1 & 1 & 1 & 1 & 1 & 1 & 0 & 0 & 0 \\
 1 & 1 & 1 & 1 & 0 & 0 & 0 & 0 & 0 & 0 & 0 & 0 & 0 & 0 & 1 & 1 & 0 \\
 -1 & -1 & -1 & -1 & 0 & 0 & 0 & 0 & -1 & -1 & -1 & -1 & -1 & -1 & -1 & -1 & 0 \\
 0 & 0 & -1 & -1 & 0 & 0 & -1 & -1 & 0 & 0 & 0 & 0 & 0 & 0 & 0 & 0 & 0 \\
 0 & 0 & 0 & 0 & 0 & -1 & 0 & -1 & 0 & 0 & 0 & 0 & -1 & -1 & 0 & 0 & 0 \\
 -1 & -1 & -1 & -1 & -1 & 0 & -1 & 0 & 0 & 0 & -1 & -1 & 0 & 0 & -1 & -1 & 0 \\
 0 & 0 & 0 & 0 & 0 & 0 & 0 & 0 & 1 & 0 & 1 & 0 & 1 & 0 & 0 & 0 & 1 \\
 1 & 0 & 1 & 0 & 0 & 0 & 0 & 0 & 0 & 1 & 0 & 1 & 0 & 1 & 1 & 0 & 0 \\
\end{smallmatrix}
\right),
\end{equation}
where we find the following $6$ non-prime factors\footnote{Note, we do not include non-prime factors related to those in the list by rotation.}
\begin{equation}
\{ p_{ 81 [23] 45 },p_{ 81 [23] 46 } ,p_{ 56 [78] 14 },p_{ 56 [78] 23 } ,p_{ 23 [45] 71 },p_{ 23 [46] 71 }\}.
\label{eq:vars_in_cone}
\end{equation}

Including these factors into the Speyer--Williams calculation we arrive at a fan with $121$ rays and maximal cones given by 
$$
 11454_8 + 1696_9 + 971_{10} + 412_{11} + 328_{12} + 89_{13} + 69_{14} + 28_{15} + 17_{16} + 7_{17},
$$
where the new `$18^{\text{th}}$' ray is given by 
\be
(1, 1, -2, 0, 0, -1, 0, 1).
\ee
An interesting question to ask is which of the maximal cones contain this new ray, they are given by
\begin{equation}
270_8+70_9+46_{10}+18_{11}+10_{12}+4_{13}+2_{14}.
\end{equation}
By searching inside the $2_{14}$ cones we are able to find the final cyclic type of quadratic non-prime factor $p_{ 56 [71] 23 }$. 

Alternatively, we can consider the subset of cones of the extended Speyer--Williams fan contained in the span of the 18 rays described above, we find $24$ such cones interestingly all of which contain the new ray; $18$ of these cones are simplicial; $4$ contain $9$ vertices; and $2$ contain $10$ vertices and are spanned by the rays
\begin{equation}
(1,2,3,4,5,6,10,15,17,18) \ \ \ \text{and}\ \ \  (6,8,9,11,12,13,14,16,17,18).
\end{equation}
Searching in these two cones we find three additional non-prime factors
\be
\{ p_{23[46]71}p_{578} - p_{23[45]71}p_{678}, p_{81[23]46}p_{578}-p_{81[23]45}p_{678},p_{56[78]23} p_{1 4 6} - p_{1 2 3} p_{4 5 6} p_{6 7 8}\},
\ee
which are the three classes of cubic $\mathcal{A}$-coordinates appearing in the ${\rm Gr}(3,8)$ cluster algebra. 

To summarise we are able to find one representative from each cyclic class of $\mathcal{A}$-coordinate appearing as non-prime factors of initial ideals of $\text{Trop}^+(I_{3,8})$ and its various extensions. Therefore, cyclic symmetry would suggest, by following the same procedure outlined above considering each of the $7$ remaining $17$-vertex cones in turn we would recover the entire set of $\mathcal{A}$-coordinates in a similar manner. After extending the ideal by the full set of $\mathcal{A}$-coordinates the results of \cite{Ilten-Najera-Treff}, applying to any finite cluster algebra of geometric type, tell us that we would again see the rays of $\text{Trop}^+(I_{3,8}^{\text{ext}})$ spanning a single maximal Gr\"obner cone whose initial ideal is generated by the forbidden pairs of the cluster algebra. 
 
 The procedure outline above for locally resolving the Speyer--Williams fan by inclusion of non-prime factors may prove useful when considering how square root letters appear for ${\rm Gr}(4,8)$ from the structure of the Gröbner fan. 
 Such square root letters were accounted for in \cite{Drummond:2019cxm} by considering coordinates associated to so called {\it limit rays} arising from limits of affine sequences of mutations inside the ${\rm Gr}(4,8)$ cluster algebra. Interestingly, these limit rays also show up in the Speyer--Williams fan, where in particular they appear in non-simplicial maximal cones with large numbers of vertices. Presumably, by attempting to locally resolve these special cones, as we have done for the case of ${\rm Gr}(3,8)$ above, we will see that instead an infinite number of non-prime factors is now needed corresponding to the infinite mutation sequences in the cluster picture. It would be interesting to study this in further detail.

\section{Missing binomials and cluster variables}

In this section we explain the techniques used above from the point of view of cluster algebras.

\subsection{The g-vectors as multiweights}\label{sec:g-vectors as valuation}

We have already seen Fomin--Zelevinsky's g-vectors appear as rays in the Speyer--Williams fan.
However, the g-vectors of Plücker coordinates or more general cluster variables carry even more information, they can detect whether a given cone in the Speyer--Williams fan corresponds to a prime cone in the positive tropicalization.
To be more precise we need to understand how g-vectors can be used as \emph{multiweights} to compute initial ideals.
For this purpose it is important to work with \emph{extended} g-vectors that can be computed recursively (just like normal g-vectors) with the only difference that also frozen nodes are taken into account. 
For example, for the quiver in Figure 3, extended g-vectors are elements in $\mathbb Z^7$ (with entries corresponding to all nodes) while normal g-vectors lie in $\mathbb Z^2$ (with entries corresponding to the active nodes $\langle13\rangle$ and $\langle14\rangle$).
Given a quiver $Q$ with $n$ active and $m$ frozen nodes we associate a matrix $B:=B_Q$ with entries for $i$ an active node and $j$ any node
\[
b_{ij}:=\#\{\text{arrows } i\to j\}- \#\{\text{arrows }j\to i\}.
\]
This matrix defines a partial order on $\mathbb Z^{n+m}$ (called the \emph{dominance order}) where $p>_Bq$ if and only if there is an element $r\in \mathbb Z_{\ge 0}^{n+m}$ such that $p=q+Br$.

\begin{Example}\label{exp}
Consider $Gr(2,4)$ with Pl\"ucker ideal $I=({p_{12}p_{34}}-p_{13}p_{24}+p{_{14}p_{23}})$.
For the cluster algebra consider the quiver
\[
\begin{tikzpicture}
\node at (0,0) {\small$\langle13\rangle$};
\node at (1.5,0) {\small $\boxed{\langle14\rangle}$};
\node at (0,-1) {\small $\boxed{\langle23\rangle}$};
\node at (1.5,-1) {\small $\boxed{\langle34\rangle}$};
\node at (-1.5,1) {\small $\boxed{\langle12\rangle}$};
\draw[->] (-1,.75) -- (-.25,.25);
\draw[->] (.375,0)-- (1,0);
\draw[->](1,-.75)--(.25,-.25);
\draw[->] (0,-.25) -- (0,-.625);
\end{tikzpicture}
\]
with associated matrix $B=(0,1,-1,-1,1)^T$ and cluster variables $s=\{p_{13},{p_{12}},{p_{14}},{p_{23}},{p_{34}}\}$. The extended g-vectors of Plücker coordinates are 
\begin{align*}
g_s(p_{13})&=(1,0,0,0,0), &
g_s(p_{24})&=(-1,0,1,1,0), &
g_s({p_{12}})&=(0,1,0,0,0),\\
g_s({p_{14}})&=(0,0,1,0,0), &
g_s({p_{23}})&=(0,0,0,1,0), &
g_s({p_{34}})&=(0,0,0,0,1).
\end{align*} 
In particular, the monomials in the Plücker relation have multiweights
\[
\overset{(0,1,0,0,1)}{{p_{12}p_{34}}} - \overset{(0,0,1,1,0)}{p_{13}p_{24}} + \overset{(0,0,1,1,0)}{{p_{14}p_{23}}}.
\]
As $(0,0,1,1,0)=(0,1,0,0,1)+(0,1,-1,-1,1)$ we deduce that $(0,0,1,1,0)>_B(0,1,0,0,1)$. Hence, the initial form is $-p_{13}p_{24}+p_{14}p_{23}$ which corresponds to a prime cone in $\text{\rm Trop}^+(I_{2,4})$.

\end{Example}

We are interested in the Gröbner fan structure of the positive part of the tropicalization. However, it is computationally expensive to compute initial ideals which is necessary to determine this fan structure. The Speyer--Williams fan (a coarsening of the g-vector fan) on the other hand is easy and fast to compute. We can use the g-vectors to compare the fan structures as follows.\footnote{This result relies on the theory of valuations and its connection to tropicalization \cite{Kaveh-Manon}. Further, it uses the fact that the g-vector fan is complete for finite type cluster algebras.}

\begin{Corollary}\label{cor}\cite{Boss_full-rank,Boss_trop-plus-cluster}
Let $s$ be a seed in the cluster algebra of a finite type Grassmannian with $n$ mutable and $m$ frozen nodes. 
Then the initial ideal of the Plücker ideal with respect to the g-vectors of Plücker coordinates obtained from $s$ is prime if and only if every point in $\mathbb Z^n\times \mathbb Z^m_{\ge 0}$ can be obtained as a positive combination of g-vectors of Plücker coordinates.
\end{Corollary}

In fact, it suffices to consider the usual g-vectors in $\mathbb Z^n$ to verify the Corollary as the extended g-vectors of frozen Plücker coordinates always span $\mathbb Z_{\ge 0}^m$.
Whenever the Corollary fails to apply to a seed $s$, the initial ideal obtained from extended g-vectors is not prime and we expect to find missing binomials.

\subsection{Cluster variables as missing binomials}
The phenomenon observed in Section~\ref{sec:Adjacency} can be explained from the point of view of cluster algebras and tropical geometry: for every cluster algebra of finite type $A$ there exists a presentation 
\begin{equation}\label{eq:presentation of A}
    A\cong \mathbb K[x_1,\dots,x_N]/I
\end{equation}
where $x_1,\dots,x_N$ correspond to \emph{all} cluster variables and $I$ is the ideal obtained by saturating\footnote{The \emph{saturation} of an ideal is an ideal that contains all elements $f$ for which there exist a monomial $x_1^{m_1}\cdots x_N^{m_N}$ such that $f\cdot x_1^{m_1}\cdots x_N^{m_N}\in I$.} the ideal generated by all exchange relations (see {\it e.g.} \cite[\S6.8]{FWZ_book}).
In the cases of the Grassmannian $\text{Gr}(3,n)$ with $n\in \{6,7,8\}$ the Pl\"ucker ideal $I_{3,n}$ is obtained from $I=I^{\text{ext}_{3,n}}$ by \emph{eliminating} those cluster variables that are not Pl\"ucker coordinates\footnote{The \emph{elimination ideal} is obtained from $I$ by intersecting with the smaller polynomial ring generated only by Pl\"ucker variables.}.

Let us assume additionally that $A$ is positively graded, as is true for the cases of interest to us.
As mentioned above \cite{Ilten-Najera-Treff} implies that there exists a unique maximal cone $C$ in the Gr\"obner fan of $I$ whose initial ideal encodes the adjacencies of all cluster variables. 
In other words, the monomial initial ideal is the Stanley--Reisner ideal of the cluster complex (cluster fan) that is minimally generated by degree two monomials of cluster variables that do not occur in any seed together (these are called \emph{non-compatible} by Fomin--Zelevinsky).
The results predict that the procedure of adding (the right) missing binomials as variables to the presentation converges.

The presentation of $A$ from \eqref{eq:presentation of A} is of such a form that is for every seed $s$ there exists a maximal cone $\tau_s\in \text{Trop}^+(I)$ such that the initial ideal with respect to the multiweights determined by g-vectors of $s$ coincides with the initial ideal of $\tau_s$ \cite{Boss_trop-plus-cluster}.
Moreover, in $I$ we have a trinomial corresponding to every exchange relation. 
It is not hard to see that for every such trinomial 
\[
f=xx'-M_1-M_2
\]
and every seed $s$ we have either
\[
\text{in}_{\tau_s}(f)=xx'-M_1 \quad \text{or} \quad \text{in}_{\tau_{s}}(f)=xx'-M_2.
\]
Suppose that one of the monomials $M_1,M_2$ consists of a single variable, say $M_1=y$. 
Then $y=xx'-M_2$ holds in $A$. All exchange relations involving $y$ yield 4-term relations in $I$. For example, 
\[
yy'=M_1'+M_2' \quad \text{gives}\quad y'(xx'-M_2) = M_1'+M_2'.
\]
What we observe in the case of the Grassmannians is that there exist weight vectors $w\in \text{Trop}^+(I)$ such that
\[
\text{in}_w(y'(xx'-M_2) - M_1' - M_2')= y'(xx'-M_2).
\]
In this case $xx'-M_2$ is a missing binomial contributing to the fact that the initial ideal is not prime.
If on the other hand $y$ is a variable in the chosen presentation of the cluster algebra then $xx'-y-M_2\in I$ and so $xx'-M_2\in \text{in}_w(I)$ is not `missing'.
This may happen precisely for those seeds (and their g-vectors) for which the Corollary~\ref{cor} does not apply.

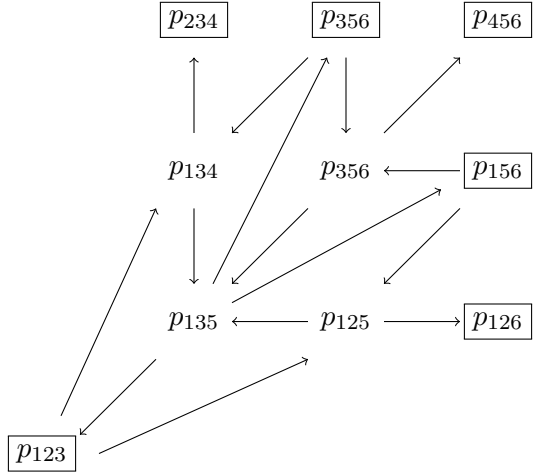
\begin{figure}
    \centering
\begin{tikzpicture}
\node at (0,0) {\small $p_{135}$};
\node at (2,0) {\small $p_{125}$};
\node at (4,0) {\small $\boxed{p_{126}}$};
\node at (0,2) {\small $p_{134}$};
\node at (2,2) {\small $p_{356}$};
\node at (4,2) {\small $\boxed{p_{156}}$};
\node at (0,4) {\small $\boxed{p_{234}}$};
\node at (2,4) {\small $\boxed{p_{356}}$};
\node at (4,4) {\small $\boxed{p_{456}}$};
\node at (-2,-1.75) {\small $\boxed{p_{123}}$};

\draw[->] (-.5,-.5) -- (-1.5,-1.5);
\draw[->] (-1.25,-1.75) -- (1.5,-.5);
\draw[->] (-1.75,-1.25) -- (-.5,1.5);
\draw[->] (1.5,0) -- (.5,0);
\draw[->] (2.5,0) -- (3.5,0);
\draw[->] (0,1.5) -- (0,.5);
\draw[->] (1.5,1.5) -- (.5,.5);
\draw[->] (.25,.5) -- (1.75,3.5);
\draw[->] (.5,.25) -- (3.25,1.75);
\draw[->] (3.5,1.5) -- (2.5,.5);
\draw[->] (3.5,2) -- (2.5,2);
\draw[->] (2.5,2.5) -- (3.5,3.5);
\draw[->] (2,3.5) -- (2,2.5);
\draw[->] (0,2.5) -- (0,3.5);
\draw[->] (1.5,3.5) -- (.5,2.5);
\end{tikzpicture}
    \caption{The $\text{Gr}(3,6)$ quiver of the seed used in Example~\ref{exp:truncated gv}.}
    \label{fig:G36}
\end{figure}

\begin{Example}\label{exp:truncated gv}
For $\text{\rm Gr}(3,6)$ take the seed $s$ with mutable cluster variables $p_{124},p_{246},p_{256}$ and $p_{346}$ (see Figure~\ref{fig:G36}). According to the computations in \S\ref{sec:Gr36} and the results in \cite{boss2021grob} this seed is associated with one of the bipyramids.
Hence, Corollary~\ref{cor} should fail in this case.
The (truncated) g-vectors of mutable Plücker coordinates with respect to $s$ are
\begin{align*}
g_s(p_{124}) &= (0,1,0,0), & g_s(p_{125}) &= (0,0,0,-1), & g_s(p_{134}) &= (-1,0,0,0),\\
g_s(p_{135}) &= (-1,-1,1,-1), & g_s(p_{136}) &= (-1,-1,1,0), & g_s(p_{145}) &= (-1,0,1,-1),\\
g_s(p_{146}) &= (-1,0,1,0), & g_s(p_{235}) &= (0,-1,1,-1), & g_s(p_{236}) &= (0,-1,1,0), \\
g_s(p_{245}) &= (0,0,1,-1), & g_s(p_{246}) &= (0,0,1,0), & g_s(p_{256}) &= (1,0,0,0),\\
g_s(p_{346}) &= (0,0,0,1), & g_s(p_{356}) &= (0,-1,0,0). &
\end{align*}
Notice that all g-vectors lie in the half space $\{(a_1,a_2,a_3,a_4)\in \mathbb Z^4:a_3\ge 0\}$. So, Corollary~\ref{cor} does not apply.
\end{Example}

\section{Massless Scattering Ideals}
\label{spinor}
So far our discussion has been focused on extracting $\mathcal{A}$-coordinates and adjacency rules from the Gr\"obner fan of the Pl\"ucker ideal for the finite type Grassmannians ${\rm Gr}(3,n)$ for $n=6,7,8$. 
In fact, in \cite{Ilten-Najera-Treff}, it was shown that the data of all cluster variables and their adjacencies in finite type cluster algebras of geometric type can be recovered from polyhedral data of a single maximal Gröbner cone of an appropriate ideal. However, what we are most interested in is how much physical information can be extracted from the Gr\"obner fan. For the cases of ${\rm Gr}(4,6)$ and ${\rm Gr}(4,7)$ the answer is the entire symbol alphabet and adjacency rules relevant for constructing the hexagon and heptagon amplitudes of planar $\mathcal{N}=4$ SYM. This motivates the question of whether the Gr\"obner fan provides a useful tool for the study of other kinematic ideals beyond the dual-conformal invariant case?

In this section we hope to provide a positive answer to this question by considering the example of general five-point massless scattering relevant for e.g. QCD processes. At two loops, all functions relevant for planar five-particle scattering were computed in  \cite{Gehrmann:2015bfy}, leading to the $26$ letter alphabet $\mathbb{A}_{p}$. This was then extended to $31$ letters relevant for the non-planar case $\mathbb{A}_{np}$ in \cite{Chicherin:2017dob} where it was used to bootstrap individual two-loop Feynman integrals. 

The goal of this section is to demonstrate how an analogous exploration of the Gr\"obner fan associated to a suitably defined five-point kinematic ideal $I_{5\text{pt}}$ can generate (almost) the entire non-planar alphabet relevant for constructing (at least at two loops) five-point massless amplitudes. Note that we do in fact miss one symbol letter, known as $W_{31}$ in \cite{Chicherin:2017dob}. However, the failure to recover $W_{31}$ is consistent with the various calculations made for five-point processes, where it has been observed to be absent from (the suitably defined finite part of) the two-loop $\mathcal{N}=4$ SYM \cite{Chicherin:2018yne,Abreu:2018aqd} and $\mathcal{N}=8$ SUGRA  \cite{Abreu:2019rpt,Chicherin:2019xeg} amplitudes at two-loops. Similar two-loop observations have been made for $q \bar{q} \rightarrow \gamma\gamma\gamma$ processes \cite{Abreu:2020cwb} and gluon amplitudes \cite{Badger:2018enw,Abreu:2019odu} in QCD.

It is important to note that the computation we present here is only an analogy to the Grassmannian cases discussed in the previous sections for two reasons. Firstly, in the case of the Grassmannian we imposed positivity conditions by considering only non-prime factors appearing in the positive part of the tropical fan $\text{Trop}^+(I_{k,n})$. However, for the five-point ideal, we do not impose any such positivity conditions and consider non-prime factors appearing in the full tropical fan $\text{Trop}(I_{5\text{pt}})$. In part this is because we are interested in non-planar theories, but also it is not totally clear, even in the planar case, which positivity conditions to impose. We suspect that there may well be multiple relevant positive regions.
Secondly, in the case of the Grassmannian, having obtained the non-prime factors, we subsequently used them in order to extend the ideal, perhaps repeating this procedure multiple times as detailed for ${\rm Gr}(3,8)$. This had the effect of eventually resolving the positive tropical fan of the fully extended ideal into a collection of simplices all with prime initial ideals. Furthermore, this singled out a single maximal Gr\"obner fan whose initial ideal contained the forbidden pairs of $\mathcal{A}$-coordinates, providing us with physical adjacency conditions on the symbol alphabet in the cases of ${\rm Gr}(4,6)$ and ${\rm Gr}(4,7)$. 
For the five-point case we perform no such extension however, again because we do not (yet) have a canonical notion of the relevant positivity conditions. Having obtained the non-prime factors appearing in the tropical fan we terminate the procedure since we already find the full non-planar alphabet. Note therefore that we do not extract any adjacency rules. That being said, it is encouraging that the same idea of symbol letters appearing as non-prime factors of the kinematic ideal follows through to the case of five-point massless scattering.
\subsection{The five-point two-loop symbol alphabet}
The kinematics of five-point massless scattering is described on the five external momenta $p_{i}^\mu$ subject to the massless on-shell condition $p_i^2=0$ and momentum conservation $\sum_{i} p_i^\mu=0$. Out of the momenta we can construct ten scalar products $s_{ij} = 2 p_i \cdot p_j$, five of which are independent upon imposing momentum conservation. 
Following the choice of \cite{Gehrmann:2018yef} they are given by\footnote{The five remaining non-adjacent scalar products may be written as $s_{13} = s_{45} -s_{12} -s_{23}$ and its cyclic rotations.}
\be
v_i = s_{ii+1} = 2 p_i \cdot p_{i+1}.
\ee
Note that there are four independent dimensionless ratios that can be formed from these variables.
It will also prove useful to introduce the following Gram determinant
\be
\Delta = \det(2 p_i \cdot p_j) = (\text{tr}_5)^2,
\ee
where we have introduced the notation $\text{tr}_5 = \text{tr}(\gamma_{5} \slashed{p}_4 \slashed{p}_5 \slashed{p}_1 \slashed{p}_2)$. Note, when written in terms of `$\beta$-variables' \cite{Bern:1993mq} $\sqrt{\Delta} = \text{tr}_5$ can be expressed as a purely rational function.

The planar (two-loop) five point alphabet $\mathbb{A}_{p} = \{  W_1,\ldots,W_{20} \} \cup \{ W_{26},\ldots,W_{31} \}$ was originally obtained in \cite{Gehrmann:2015bfy} and consists of $26$ letters given by 
\begin{align}
&W_i=v_i, & &W_{5+i}=v_{i+2}+v_{i+3}, & &W_{10+i}=v_{i}-v_{i+3}, \notag\\
&W_{15+i}= v_{i+3}-v_i -v_{i+1}, & &W_{25+i}=\frac{a_{i } - \sqrt{\Delta}}{ a_{i }+ \sqrt{\Delta}}, & &W_{31}= \sqrt{\Delta},
\end{align}
where the $i$ indices run from $1$ to $5$ and we have introduced the notation $$a_{i} = v_{i} v_{i+1} -v_{i+1} v_{i+2} +v_{i+2} v_{i+3} -v_{i} v_{i+4} -v_{i+3} v_{i+4}.$$ By closing the planar alphabet under permutations the authors of \cite{Chicherin:2017dob} generalised the alphabet to the non-planar case $\mathbb{A}_{np} = \mathbb{A}_p \cup \{ W_{21},\ldots W_{26} \}$, where we introduce the five additional non-planar letters given by
\be
W_{20 +i }=v_{i+2} +v_{i+3}-v_{i}-v_{i+1}.
\ee
As for the planar $\mathcal{N}=4$ SYM case the non planar alphabet $\mathbb{A}_{np}$ provides the starting point for the construction of integrable polylogarithmic symbols relevant for the bootstrap of five-point massless non-planar amplitudes/integrals \cite{Chicherin:2017dob}. 
\subsection{Non-planar alphabet from a Gr\"obner fan}
Inspired by the appearance of symbol letters ($\mathcal{A}$-coordinates) as non-prime factors of the Pl\"ucker ideal we wish to apply similar ideas to a suitbaly defined five-point ideal $I_{5\text{pt}}$ in order to generate the non-planar alphabet $\mathbb{A}_{np}$.

To define the kinematic space for general $n$-point massless scattering, instead of using momentum twistor variables, it is instructive to consider spinor-helicity variables. We introduce these in the usual way by defining each null momentum as a bispinor,
\be
p_i^{\alpha \dot{\alpha}} = (\sigma_\mu)^{\alpha \dot{\alpha}} p_i^\mu = \lambda_i^\alpha \tilde{\lambda}_i^{\dot \alpha}\,.
\ee
Then we consider the Lorentz invariant brackets\footnote{Note that we are interested in the generic case without dual conformal symmetry. One could use momentum twistors also in this case, allowing for the appearance of the infinity bitwistor, i.e. for both four-brackets $\langle ijkl \rangle$ and two-brackets $\langle ij \rangle = \langle ij I \rangle$, however these variables are most natural in the case of planar amplitudes with ordered external legs. Here we will not impose any particular ordering.}
\be
\langle ij \rangle = \lambda_i^\alpha \lambda_{j \alpha}\,, \qquad [ij] = \tilde{\lambda}_i^{\dot{\alpha}} \tilde{\lambda}_{i \dot{\alpha}}\,.
\ee
The spinor brackets defined above are constrained by two sets of relations, the Pl\"ucker (or Schouten) identities, 
\begin{align}
\langle ij \rangle \langle kl \rangle -\langle ik \rangle \langle jl \rangle +\langle il \rangle \langle jk \rangle=0; \quad  [ ij ] [ kl ] -[ ik ] [ jl ] +[ il ] [ jk ]=0, 
\label{SH_ideal1}
\end{align}
for $1\leq i<j<k<l \leq n$ and momentum conservation,
\be
\sum_{j=1}^n \langle ij \rangle [jk] =0,
\label{SH_ideal2}
\ee
for $i,k \in \{1,\ldots,n\}$. Note that the Pl\"ucker relations are three-term relations, while the momentum conservation relations can be three-term (for the off-diagonal case $i\neq k$) or four-term (for the diagonal case $i=k$). 

As we wish to recast the kinematic space in the language of polynomial ideals, we will introduce new variables $a_{ij}$ and $\tilde{a}_{ij}$ which obey the corresponding polynomial relations,
\begin{align}
a_{ij} a_{kl} -a_{ik} a_{jl} +a_{il}a_{jk} =0\,, \quad  \tilde{a}_{ij} \tilde{a}_{kl} -\tilde{a}_{ik} \tilde{a}_{jl} +\tilde{a}_{il}\tilde{a}_{jk} =0\,, \quad \sum_{j=1}^n a_{ij} \tilde{a}_{jk} =0\,.
\label{SH_ideal3}
\end{align}
We take these relations to define an ideal $I_{n {\rm pt}}$ on $2 \times {n \choose 2}$ variables $a_{ij}$ and $\tilde{a}_{ij}$.

We will study the ideal, $I_{5 {\rm pt}}$, defined on the ten $a_{ij}$ and ten $\tilde{a}_{ij}$ variables with $i<j$ organised as $(a_{12},\ldots,a_{45},\tilde{a}_{12},\ldots,\tilde{a}_{45})$ and subject to (\ref{SH_ideal3}) for $n=5$. In particular, we focus our attention on the  Gr\"obner fan $GF(I_{5 {\rm pt}})$ and tropical fan $\text{Trop}(I_{5 {\rm pt}})$. Note the ideal has a six-dimensional linear subspace (or lineality space), corresponding to the five little group rescalings of the spinor variables,
\be
\lambda_i \rightarrow \alpha_i \lambda_i,\qquad \tilde{\lambda}_i \rightarrow \alpha_i^{-1} \tilde{\lambda}_i\,; \qquad a_{ij} \rightarrow \alpha_i \alpha_j a_{ij}\,, \qquad \tilde{a}_{ij} \rightarrow \alpha_i^{-1} \alpha_j^{-1} \tilde{a}_{ij}\,.
\ee
and the single overall dimension rescaling,
\be
\lambda_i \rightarrow \beta \lambda_i \,, \qquad \tilde{\lambda}_i \rightarrow \beta \tilde{\lambda_i}\,; \qquad a_{ij} \rightarrow \beta^2 a_{ij}\,, \qquad \tilde{a}_{ij} \rightarrow \beta^2 \tilde{a}_{ij}\,.
\ee
In fact we can identify the ideal $I_{5 {\rm pt}}$ with the Grassmannian Pl\"ucker ideal $I_{3,6}$ via the following identification of variables,
\be
a_{ij} = p_{ij6}\,, \qquad \tilde{a}_{ij} = (-1)^{j-i-1}p_{klm}\,,
\ee
where in the second equality $\{k,l,m\}$ is the ordered complement of $\{i,j\}$ in the set $\{1,2,3,4,5\}$. Thus the $a_{ij}$ correspond to all the Pl\"ucker variables involving the label $6$, while the $\tilde{a}_{ij}$ give all those which do not. The Pl\"ucker relations for the $a_{ij}$ and the $\tilde{a}_{ij}$ together with the off-diagonal three-term momentum conservation relations, give the three-term Pl\"ucker relations for ${\rm Gr}(3,6)$ while the five diagonal four-term momentum conservation relations are the minimal set of four-term Pl\"ucker relations needed to complete a generating set of the Pl\"ucker ideal for ${\rm Gr}(3,6)$. 

Thus to study the tropical fan ${\rm Trop}(I_{5 {\rm pt}})$ we can simply refer to results already described for ${\rm Gr}(3,6)$. In particular we know the fan has the f-vector (\ref{fvectTropGr36}) and, in particular, has $65$ rays. In the variables $a_{ij}$ and $\tilde{a}_{ij}$ used to define $I_{5pt}$ these rays take the following form: $20$ unit vectors ${\bf e}_{ij}$ and $\tilde{{\bf e}}_{ij}$ where ${\bf e}_{ij}$ is the unit vector in the $a_{ij}$ direction and respectively $\tilde{{\bf e}}_{ij}$ is the unit vector in the $\tilde{a}_{ij}$ direction; $5$ vectors of the form ${\bf z}_i = \sum_{j \neq i} {\bf e}_{ij}$, lineality equivalent to  $\tilde{{\bf z}}_i = \sum_{j \neq i} \tilde{{\bf e}}_{ij}$; 10 permutations of ${\bf r}_{45} = {\bf v}_{123} + \tilde{{\bf e}}_{45}$, lineality equivalent to $\tilde{{\bf r}}_{45} = \tilde{{\bf v}}_{123} + {\bf e}_{45}$, where we have defined ${\bf v}_{ijk} = {\bf e}_{ij}+{\bf e}_{ik}+{\bf e}_{jk}$ and similarly for $\tilde{{\bf v}}_{ijk}$; finally the $30$ permutations of ${\bf y}_{1,23,45} = {\bf v}_{123} + \tilde{{\bf v}}_{145} = \tilde{\bf y}_{1,23,45}$.

\begin{figure}[h]
\centering
\includegraphics[width=\linewidth]{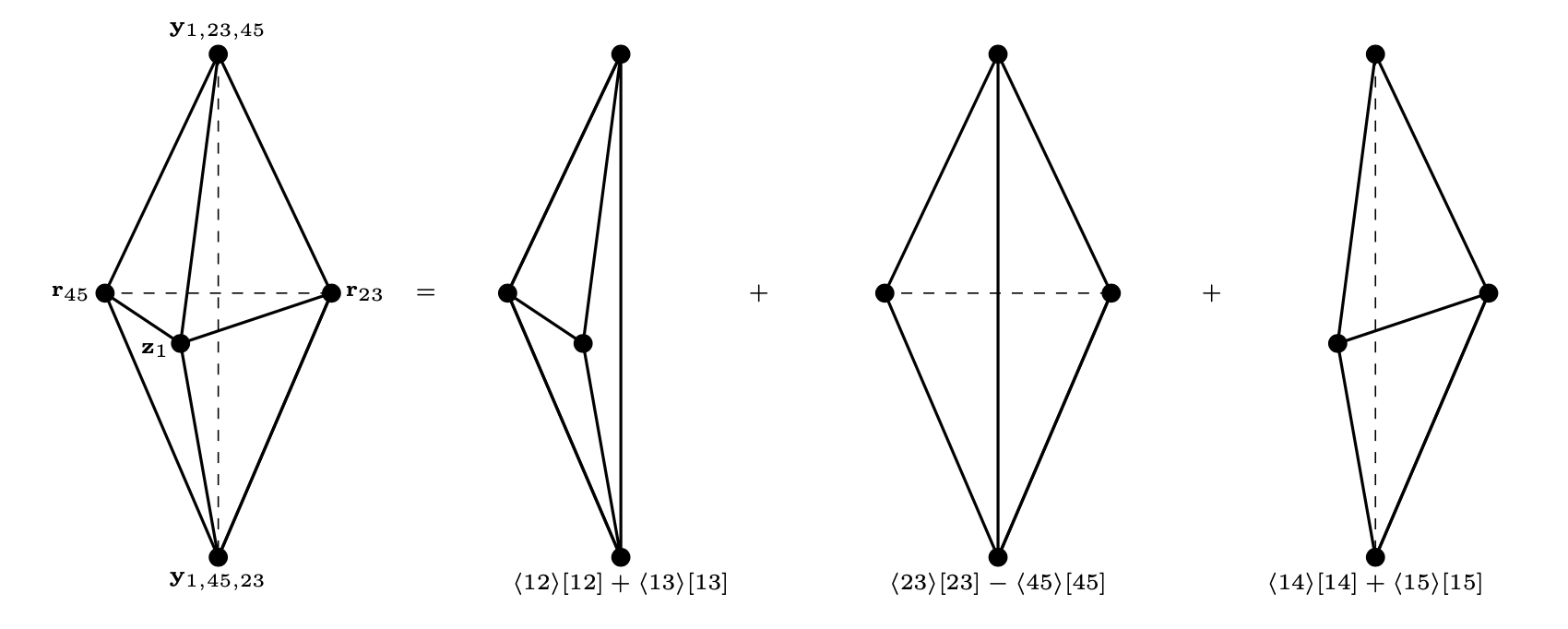}
\caption{\small A bipyramid $b_1$ of $\text{Trop}(I_{5 {\rm pt}})$, on the left hand side we have the full bipyramid labelled by its $5$ rays, on the right hand side the bipyramid is split into three tetrahedra. Each tetrahedron is labelled by the quadratic non-prime factor found in the initial ideal, where all three expressions are equivalent modulo $I_{5\text{pt}}$.}
\label{fig:grob_split_5}
\end{figure}

Amongst the $1035$ maximal cones only $45$ are non-prime: $30$ given by the permutation copies of $\{ {\bf y}_{1,23,45},{\bf z}_1,{\bf r}_{45}, {\bf y}_{1,45,23} \}$; and an additional $15$ given by the permutation copies of $\{ {\bf y}_{1,23,45},{\bf r}_{23},{\bf r}_{45}, {\bf y}_{1,45,23} \}$. As already described in our discussion of ${\rm Gr}(3,6)$, these tetrahedra fit together into bipyramids. In fat we have $15$ such bipyramids across the full tropical fan.  As an example consider Figure \ref{fig:grob_split_5} where we have a single bipyramid: the left tetrahedron with the rays 
$
\{ {\bf y}_{1,23,45},{\bf z}_1,{\bf r}_{45}, {\bf y}_{1,45,23} \},
$
produces the non-prime factor $a_{12} \tilde{a}_{12} + a_{13} \tilde{a}_{13} = \langle 12 \rangle [12] + \langle 13 \rangle [13] $; transposing $2 \leftrightarrow 4$ and $3 \leftrightarrow 5$ we find the right tetrahedron with the rays 
$
\{ {\bf y}_{1,23,45},{\bf z}_1,{\bf r}_{23}, {\bf y}_{1,45,23} \},
$
and non-prime factor  $a_{14} \tilde{a}_{14} + a_{15} \tilde{a}_{15} = \langle 14 \rangle [14] + \langle 15 \rangle [15]$; finally the middle tetrahedron has the rays given by 
$
\{ {\bf y}_{1,23,45},{\bf r}_{23},{\bf r}_{45}, {\bf y}_{1,45,23} \},
$
which produces the non-prime factor $a_{23} \tilde{a}_{23} - a_{45} \tilde{a}_{45} = \langle 23 \rangle [23] - \langle 45 \rangle [45]$. The three non-prime factors appearing in the bipyramid are equivalant modulo the ideal $I_{5 \text{pt}}$.

To generate the full set of non-prime factors modulo the ideal $I_{5\text{pt}}$ we need only take the permutation copies of $\langle 23 \rangle [23] - \langle 45 \rangle [45]$ which produces $15$  quadratic expressions given by
\begin{align}
& \langle 23 \rangle [23] - \langle 45 \rangle [45], &&   \langle 24 \rangle  [24] - \langle 35 \rangle [35], &&  \langle 25 \rangle [25] - \langle 34 \rangle [34], \notag \\
& \langle 13 \rangle [13] - \langle 45 \rangle [45], &&   \langle 14 \rangle [14] - \langle 35 \rangle [35], &&  \langle 15 \rangle [15] - \langle 34 \rangle [34],\notag \\
& \langle 12 \rangle [12] - \langle 45 \rangle [45], &&   \langle 14 \rangle [14] - \langle 25 \rangle [25], &&  \langle 15 \rangle [15] - \langle 24 \rangle [24],\notag \\
& \langle 12 \rangle [12] - \langle 35 \rangle [35], &&   \langle 13 \rangle [13] - \langle 25 \rangle [25], &&  \langle 15 \rangle [15] - \langle 23 \rangle [23],\notag \\
& \langle 12 \rangle [12] - \langle 34 \rangle [34], &&   \langle 13 \rangle [13] - \langle 24 \rangle [24], &&  \langle 14 \rangle [14] - \langle 23 \rangle [23].
\label{non_prime_15}
\end{align}
 Along with the $\{a_{ij} = \langle ij \rangle, \tilde{a}_{ij} = [ij] \}$ this provides us $35$ expressions from which to form homogenous combinations.

To see that we are in fact recovering the same content as the symbol alphabet $\mathbb{A}_{np}$ we re-write the entire non-planar alphabet in terms of spinor helicity variables given by the cyclic copies of\footnote{Let us emphasise the fact that in spinor-helicity variables the rationality of letters $\{ W_{26}, \ldots, W_{31}\}$ becomes manifest.}
\begin{align}
&W_1= \langle 12 \rangle [12] , & &W_{6}=   \langle 34 \rangle [34]+ \langle 45 \rangle [45], \notag \\
&W_{11}=\langle 34 \rangle [34]+\langle 35 \rangle [35], && W_{16}=\langle 13 \rangle [13], \notag \\
&W_{21} = \langle 13 \rangle [13] + \langle 34 \rangle [34], & &W_{26}=\frac{\langle 45 \rangle [51] \langle  12 \rangle [24]}{ [45] \langle 51 \rangle  [12] \langle 24 \rangle}, \notag \\
& W_{31}= [45] \langle 51 \rangle  [12] \langle 24 \rangle-\langle 45 \rangle [51] \langle  12 \rangle [24]. && 
\end{align}
With this representation it is clear that letters $\{W_i \}_{i=1}^5 \cup \{W_i \}_{i=16}^{20} \cup \{W_i \}_{i=26}^{30} $ are given by multiplicative combinations of the $\{a_{ij},\tilde{a}_{ij} \}$ variables. Furthermore, the remaining $15$ letters $\{ W_{i} \}_{i=6}^{10} \cup \{ W_{i} \}_{i=11}^{15}  \cup \{ W_{i} \}_{i=21}^{25}$, themselves related by the $\mathcal{S}_5$ permutation symmetry, are exactly the $15$ non-prime factors appearing in the Gr\"obner fan of the spinor-helicity ideal! To see this explicitly note we have
\begin{align}
& W_{6\ } =  \langle 34 \rangle [34]+ \langle 45 \rangle [45] =  \langle 12 \rangle [12] - \langle 35 \rangle [35], \notag \\
& W_{11} =  \langle 34 \rangle [34]+ \langle 35 \rangle [35] =  \langle 12 \rangle [12] - \langle 45 \rangle [45], \notag \\
& W_{21} =  \langle 13 \rangle [13]+ \langle 34 \rangle [34] =  \langle 14 \rangle [14] - \langle 25 \rangle [25],
\end{align}
all of which appear in \eqref{non_prime_15}. It follows then that taking homogenous combinations of letters $\{ W_{i }\}_{i=1}^{30}$ is equivalent to taking homogenous combinations of $\{ \langle ij \rangle, [ij] \}$ and the permutations of the non-prime factors $\langle 23 \rangle [23] - \langle 45 \rangle [45]$. Note, as already emphasised, we do not recover the letter $W_{31}$. However, this is consistent with $W_{31}$ not appearing in the expressions for suitably defined amplitudes.



\section{Conclusion}
We have presented a prescription for exploring the {\it alphabet} associated to the kinematic space of scattering amplitudes via the Gr\"obner fan. The construction was demonstrated for the cases ${\rm Gr}(3,n)$ with $n=6,7,8$ to extract the $\mathcal{A}$-coordinates from a combination of the Speyer--Williams and Gr\"obner fans. By searching for non-prime factors appearing in initial ideals associated to cones within the positive part of the tropical fan we are able to recover the full set of $\mathcal{A}$-coordinates. Having obtained the full set of $\mathcal{A}$-coordinates, following the results of \cite{boss2021grob}, we are able to extract the forbidden pairs of $\mathcal{A}$-coordinates by computing the initial ideal of a Gröbner cone singled out by the positive tropical part of the extended ideal. In the case of ${\rm Gr}(3,7)$ this corresponds to the alphabet and adjacency rules relevant for constructing seven point amplitudes in planar $\mathcal{N}=4$ SYM. As a further application of these techniques it would be interesting to consider ${\rm Gr}(4,8)$ where the corresponding cluster algebra becomes infinite. We should expect to see infinite sequences of non-prime factors appearing when resolving maximal cones containing {\it limit rays}.

We have also provided hints that the Gr\"obner fan may be a useful tool when considering more general scattering processes. We were able to find the entire non-planar alphabet relevant for five-point non dual conformal scattering processes by considering the Gr\"obner fan of a suitably defined kinematic ideal. In fact, this ideal turns out to be none other than the Grassmannian Pl\"ucker ideal for ${\rm Gr}(3,6)$. Note, as emphasised in the main text no positivity criteria was imposed in this calculation. It would be interesting to study this case further to determine what is the correct notion of positivity for reducing to the planar alphabet for five-point scattering (clearly this will not be the same as reducing to the positive region in ${\rm Gr}(3,6)$) and whether this sheds light on any adjacency properties. 

Finally, the six-point alphabet was recently studied at one-loop by \cite{Henn:2022ydo}, it would be interesting to apply the techniques developed here to see how much of the alphabet we can obtain.

\appendix
\section{Details of ${\rm Trop}^+(I_{3,7})$}\label{app:Gr37}\setcounter{equation}{0}

Let us look at the $7$-vertex cones in the Speyer--Williams fan for ${\rm Gr}(3,7)$. There are $63$ cones with $7$ vertices which come in five dihedral orbits each of length $\{14, 14, 7, 14, 14\} $. They have representatives with rays given by 
\be
\left(
\begin{array}{ccccccc}
 b_{\text{4,2345671}} & b_{\text{3,2345671}} & b_{\text{2,7123456}} & b_{\text{5,2345716}} & b_{\text{4,2371456}} & b_{\text{4,4567123}} & b_{\text{3,4567123}} \\
 b_{\text{5,7123564}} & b_{\text{1,1234567}} & b_{\text{4,7123456}} & b_{\text{3,2345671}} & b_{\text{4,5671234}} & b_{\text{2,5671234}} & b_{\text{3,5671234}} \\
 b_{\text{5,7123564}} & b_{\text{1,1234567}} & b_{\text{4,7123456}} & b_{\text{5,5671342}} & b_{\text{2,5671234}} & b_{\text{4,5634712}} & b_{\text{1,3456712}} \\
 b_{\text{5,7123564}} & b_{\text{4,7123456}} & b_{\text{4,2345671}} & b_{\text{2,5671234}} & b_{\text{4,7156234}} & b_{\text{1,3456712}} & b_{\text{3,7123456}} \\
 b_{\text{1,1234567}} & b_{\text{3,2345671}} & b_{\text{4,4567123}} & b_{\text{3,4567123}} & b_{\text{4,6745123}} & b_{\text{3,6712345}} & b_{\text{2,4567123}} \\
\end{array}
\right)
\label{7conereps}
\ee
Starting from top to bottom the cones are refined as 
\be
\left(
\begin{array}{cccccc}
 b_{\text{2,7123456}} & b_{\text{3,4567123}} & b_{\text{3,2345671}} & b_{\text{4,2345671}} & b_{\text{4,2371456}} & b_{\text{4,4567123}} \\
 b_{\text{2,7123456}} & b_{\text{3,4567123}} & b_{\text{4,2345671}} & b_{\text{4,2371456}} & b_{\text{4,4567123}} & b_{\text{5,2345716}} \\
\end{array}
\right),
\ee
containing the missing binomials $p_{145} p_{237} - p_{123} p_{457}$ and $p_{137} p_{245} - p_{127} p_{345}$ respectively for the cone defined by the first row. We then have
\be
\left(
\begin{array}{cccccc}
 b_{\text{1,1234567}} & b_{\text{2,5671234}} & b_{\text{3,2345671}} & b_{\text{4,7123456}} & b_{\text{4,5671234}} & b_{\text{5,7123564}} \\
 b_{\text{1,1234567}} & b_{\text{3,5671234}} & b_{\text{3,2345671}} & b_{\text{4,7123456}} & b_{\text{4,5671234}} & b_{\text{5,7123564}} \\
 b_{\text{1,1234567}} & b_{\text{2,5671234}} & b_{\text{3,5671234}} & b_{\text{3,2345671}} & b_{\text{4,5671234}} & b_{\text{5,7123564}} \\
\end{array}
\right),
\ee
for the second row containing the missing binomials $\{ p_{167} p_{345} - p_{157} p_{346}, p_{167} p_{245} - p_{157} p_{246} \}$, $\{ p_{147} p_{356} - p_{137} p_{456}, p_{147} p_{256} - p_{127} p_{456} \}$ and $\{p_{156} p_{347} - p_{134} p_{567}, p_{156} p_{247} - p_{124} p_{567} \}$ respectively. Note that $p_{167} p_{245} - p_{157} p_{246}$ is in a different cyclic class to the missing binomials found up to this point. Thus it is already clear at this stage that we will find all 14 quadratic $\mathcal{A}$-coordinates appearing as factors of non-prime initial ideals. For the third row we have
\be
\left(
\begin{array}{cccccc}
 b_{\text{1,3456712}} & b_{\text{1,1234567}} & b_{\text{2,5671234}} & b_{\text{4,7123456}} & b_{\text{5,7123564}} & b_{\text{5,5671342}} \\
 b_{\text{1,3456712}} & b_{\text{1,1234567}} & b_{\text{4,7123456}} & b_{\text{4,5634712}} & b_{\text{5,7123564}} & b_{\text{5,5671342}} \\
 b_{\text{1,3456712}} & b_{\text{1,1234567}} & b_{\text{2,5671234}} & b_{\text{4,5634712}} & b_{\text{5,7123564}} & b_{\text{5,5671342}} \\
\end{array}
\right)\,,
\ee
 with the missing binomials $p_{167} p_{245} - p_{157} p_{246}$, $ p_{147} p_{256} - p_{127} p_{456}$ and $p_{156} p_{247} - p_{124} p_{567}$. For the fourth row we have
\be
\left(
\begin{array}{cccccc}
 b_{\text{1,3456712}} & b_{\text{2,5671234}} & b_{\text{4,2345671}} & b_{\text{4,7156234}} & b_{\text{4,7123456}} & b_{\text{5,7123564}} \\
 b_{\text{1,3456712}} & b_{\text{2,5671234}} & b_{\text{3,7123456}} & b_{\text{4,2345671}} & b_{\text{4,7156234}} & b_{\text{4,7123456}} \\
\end{array}
\right)\,,
\ee
containing the missing binomials $p_{156} p_{237} - p_{123} p_{567}$ and $p_{167} p_{235} - p_{157} p_{236}$.

Before moving onto the final row of (\ref{7conereps}) let us observe here that none of the above refinements has utilised an extra positive ray of $b_6$ type. The remaining case does however. For convenience we repeat the seven vertices here,
\be
\left\{ \textcolor{DarkGreen}{b_{\text{1,1234567}} }, \textcolor{DarkGreen}{b_{\text{4,4567123}}}, \textcolor{DarkGreen}{b_{\text{4,6745123}}}, \textcolor{red}{b_{\text{3,2345671}}}  , \textcolor{blue}{b_{\text{3,4567123}}}  , \textcolor{pink}{b_{\text{3,6712345}}} , \textcolor{yellow}{b_{\text{2,4567123}}} \right\}\,.
\ee
This 7-cone is refined as follows,
\be
\left(
\begin{array}{cccccc}
 \textcolor{DarkGreen}{b_{\text{1,1234567}}} &  \textcolor{DarkGreen}{b_{\text{4,4567123}}}  & \textcolor{DarkGreen}{b_{\text{4,6745123}}} & \textcolor{yellow}{b_{\text{2,4567123}}} & \textcolor{red}{b_{\text{3,2345671}}} & \textcolor{blue}{b_{\text{3,4567123}}}  \\
  \textcolor{DarkGreen}{b_{\text{1,1234567}}} &  \textcolor{DarkGreen}{b_{\text{4,4567123}}}  & \textcolor{DarkGreen}{b_{\text{4,6745123}}} & \textcolor{yellow}{b_{\text{2,4567123}}} & \textcolor{red}{b_{\text{3,2345671}}} & \textcolor{pink}{b_{\text{3,6712345}}} \\
   \textcolor{DarkGreen}{b_{\text{1,1234567}}} &  \textcolor{DarkGreen}{b_{\text{4,4567123}}}  & \textcolor{DarkGreen}{b_{\text{4,6745123}}}  & \textcolor{orange}{b_{\text{6,4567123}}} & \textcolor{red}{b_{\text{3,2345671}}}  &  \textcolor{blue}{b_{\text{3,4567123}}}    \\
   \textcolor{DarkGreen}{b_{\text{1,1234567}}} &  \textcolor{DarkGreen}{b_{\text{4,4567123}}}  & \textcolor{DarkGreen}{b_{\text{4,6745123}}} & \textcolor{orange}{b_{\text{6,4567123}}}&\textcolor{red}{b_{\text{3,2345671}}} & \textcolor{pink}{b_{\text{3,6712345}}}     \\
  \textcolor{DarkGreen}{b_{\text{1,1234567}}} &  \textcolor{DarkGreen}{b_{\text{4,4567123}}}  & \textcolor{DarkGreen}{b_{\text{4,6745123}}} & \textcolor{orange}{b_{\text{6,4567123}}} &  \textcolor{pink}{b_{\text{3,6712345}}}  &  \textcolor{blue}{b_{\text{3,4567123}}}   \\
\end{array}
\right)\,,
\ee
with each part providing the following missing binomials
\be
\left(
\begin{matrix}
p_{237} p_{456} - p_{236} p_{457},& p_{137} p_{456} - p_{136} p_{457},& p_{127} p_{456} - p_{126} p_{457}\\
p_{235} p_{467} - p_{234} p_{567},& p_{135} p_{467} - p_{134} p_{567},& p_{125} p_{467} - p_{124} p_{567}\\
p_{237} p_{456} - p_{236} p_{457},& p_{167} p_{345} - p_{145} p_{367},& p_{167} p_{245} - p_{145} p_{267}\\
p_{235} p_{467} - p_{234} p_{567},& p_{167} p_{345} - p_{145} p_{367},& p_{167} p_{245} - p_{145} p_{267}\\
p_{267} p_{345} - p_{245} p_{367},& p_{167} p_{345} - p_{145} p_{367},& p_{167} p_{245} - p_{145} p_{267}
\end{matrix}
\right)\,.
\ee
We illustrate this refinement below in Figure \ref{fig:7vertexsplit}.

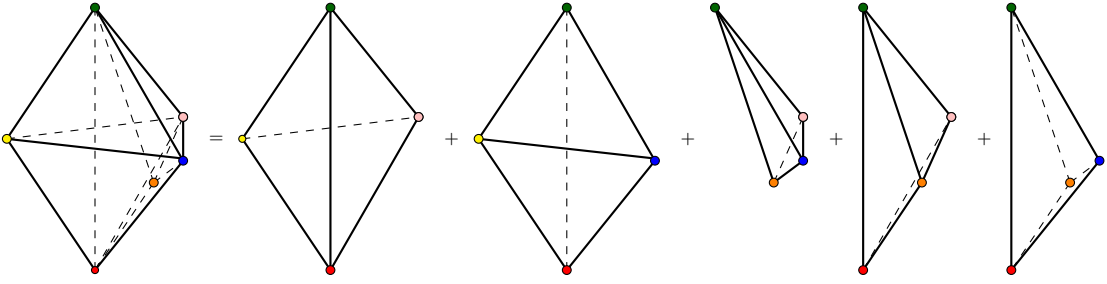
\begin{figure}[h]
\centering
\begin{tikzpicture}[scale=0.58]
       
    \draw[dashed] (2,3) -- (2,-3);
     \draw[thick] (2,3) -- (4,-0.5);
     \draw[thick] (2,-3) -- (4,-0.5);
     \draw[thick] (2,3) -- (4,0.5);
     \draw[dashed] (2,-3) -- (4,0.5);
     \draw[thick]  (4,-0.5) -- (4,0.5);
     \draw[thick]  (0,0) -- (4,0.-0.45);
     \draw[dashed]  (0,0) -- (4,0.5);
      \draw[thick]  (0,0) -- (2,-3);
       \draw[thick]  (0,0) -- (2,3);
      \draw[dashed]  (3.3333,-1) -- (2,-3);
       \draw[dashed]  (3.3333,-1) -- (4,-0.5);
       \draw[dashed]  (3.3333,-1) -- (4,0.5);
       \draw[dashed]  (3.3333,-1) -- (2,3);

    \draw[fill=DarkGreen] (2,3) circle (0.1);
     \draw[fill=red] (2,-3) circle (0.08);
     \draw[fill=blue] (4,-0.5) circle (0.1);
    \draw[fill=pink] (4,0.5) circle (0.1);
    \draw[fill=pink] (4,0.5) circle (0.1);      
    \draw[fill=yellow] (0,0) circle (0.1);
    \draw[fill=orange] (3.3333,-1) circle (0.1);
    \draw[fill=white] (5,0) circle (0.00);  
    \draw[fill=white] (2,-4) circle (0.00);    
      \draw[thick] (4.75,0) node {{\tiny $ = $}};

 \end{tikzpicture}
  \begin{tikzpicture}[scale=0.58]
       
    \draw[thick] (2,3) -- (2,-3);
     \draw[thick] (2,3) -- (4,0.5);
     \draw[thick] (2,-3) -- (4,0.5);
     \draw[dashed]  (0,0) -- (4,0.5);
      \draw[thick]  (0,0) -- (2,-3);
       \draw[thick]  (0,0) -- (2,3);

    \draw[fill=DarkGreen] (2,3) circle (0.1);
     \draw[fill=red] (2,-3) circle (0.1);
     \draw[fill=blue] (4,-0.5) circle (0.00);
    \draw[fill=pink] (4,0.5) circle (0.1);
    \draw[fill=pink] (4,0.5) circle (0.1);      
    \draw[fill=yellow] (0,0) circle (0.08);
    \draw[fill=yellow] (3.3333,-1) circle (0.00);
    \draw[fill=white] (5,0) circle (0.00);  
    \draw[fill=white] (2,-4) circle (0.00);    
      \draw[thick] (4.75,0) node {{\tiny $ +$}};


 \end{tikzpicture}
 \begin{tikzpicture}[scale=0.58]
       
    \draw[dashed] (2,3) -- (2,-3);
     \draw[thick] (2,3) -- (4,-0.5);
     \draw[thick] (2,-3) -- (4,-0.5);
     \draw[thick]  (0,0) -- (4,0.-0.45);
      \draw[thick]  (0,0) -- (2,-3);
       \draw[thick]  (0,0) -- (2,3);

    \draw[fill=DarkGreen] (2,3) circle (0.1);
     \draw[fill=red] (2,-3) circle (0.1);
     \draw[fill=blue] (4,-0.5) circle (0.1);
    \draw[fill=pink] (4,0.5) circle (0.00);   
    \draw[fill=yellow] (0,0) circle (0.1);
    \draw[fill=orange] (3.3333,-1) circle (0.00);
    \draw[fill=white] (5,0) circle (0.00);  
    \draw[fill=white] (2,-4) circle (0.00);    
      \draw[thick] (4.75,0) node {{\tiny $ +$}};

 \end{tikzpicture}
 \begin{tikzpicture}[scale=0.58]
       
     \draw[thick] (2,3) -- (4,-0.5);
     \draw[thick] (2,3) -- (4,0.5);
     \draw[thick]  (4,-0.5) -- (4,0.5);
       \draw[thick]  (3.3333,-1) -- (4,-0.5);
       \draw[dashed]  (3.3333,-1) -- (4,0.5);
       \draw[thick]  (3.3333,-1) -- (2,3);

    \draw[fill=DarkGreen] (2,3) circle (0.1);
     \draw[fill=white] (2,-3) circle (0.00);
     \draw[fill=blue] (4,-0.5) circle (0.1);
    \draw[fill=pink] (4,0.5) circle (0.1);
    \draw[fill=pink] (4,0.5) circle (0.1);      
    \draw[fill=orange] (3.3333,-1) circle (0.1);
    \draw[fill=white] (5,0) circle (0.00);  
    \draw[fill=white] (2,-4) circle (0.00);    
      \draw[thick] (4.75,0) node {{\tiny $ +$}};

 \end{tikzpicture}
 \begin{tikzpicture}[scale=0.58]
       
    \draw[thick] (2,3) -- (2,-3);
     \draw[thick] (2,3) -- (4,0.5);
     \draw[dashed] (2,-3) -- (4,0.5);
      \draw[thick]  (3.3333,-1) -- (2,-3);
       \draw[thick]  (3.3333,-1) -- (4,0.5);
       \draw[thick]  (3.3333,-1) -- (2,3);

    \draw[fill=DarkGreen] (2,3) circle (0.1);
     \draw[fill=red] (2,-3) circle (0.1);
    \draw[fill=pink] (4,0.5) circle (0.1);
    \draw[fill=pink] (4,0.5) circle (0.1);      
    \draw[fill=orange] (3.3333,-1) circle (0.1);
    \draw[fill=white] (5,0) circle (0.00);  
    \draw[fill=white] (2,-4) circle (0.00);    
      \draw[thick] (4.75,0) node {{\tiny $ +$}};

 \end{tikzpicture}
 \begin{tikzpicture}[scale=0.58]
       
    \draw[thick] (2,3) -- (2,-3);
     \draw[thick] (2,3) -- (4,-0.5);
     \draw[thick] (2,-3) -- (4,-0.5);
      \draw[dashed]  (3.3333,-1) -- (2,-3);
       \draw[dashed]  (3.3333,-1) -- (4,-0.5);
       \draw[dashed]  (3.3333,-1) -- (2,3);

    \draw[fill=DarkGreen] (2,3) circle (0.1);
     \draw[fill=red] (2,-3) circle (0.1);
     \draw[fill=blue] (4,-0.5) circle (0.1);
    \draw[fill=orange] (3.3333,-1) circle (0.1);
    \draw[fill=white] (5,0) circle (0.00);  
    \draw[fill=white] (2,-4) circle (0.00);    
      \draw[thick] (4.75,0) node {{\tiny $ $}};

 \end{tikzpicture}
    
\caption{\small A 7-vertex Speyer--Williams cone is refined into $5$ simplices in the tropical fan, three of which use the spurious orange vertex and two of which do not.}
\label{fig:7vertexsplit}
\end{figure}
Next we consider the $8$-vertex cones of the Speyer--Williams fan. There are $28$ of these coming in two dihedral classes. The first has a representative given by 
\be
\left\{b_{\text{4,7123456}},b_{\text{4,2345671}},b_{\text{3,2345671}},b_{\text{2,7123456}},b_{\text{5,2345716}},b_{\text{4,2371456}},b_{\text{3,4567123}},b_{\text{3,7123456}}\right\}\,,
\ee
which has the refinement
\be
\left(
\begin{smallmatrix}
 b_{\text{2,7123456}} & b_{\text{3,4567123}} & b_{\text{3,2345671}} & {b_{\text{4,2371456}}} & {b_{\text{4,7123456}}} &  b_{\text{4,2345671}} \\
 b_{\text{2,7123456}} & b_{\text{3,4567123}} & b_{\text{4,2345671}} & {b_{\text{4,2371456}}} & {b_{\text{4,7123456}}}& b_{\text{5,2345716}} \\
 b_{\text{3,4567123}} & b_{\text{3,7123456}} & b_{\text{4,2345671}} & {b_{\text{4,2371456}}} & {b_{\text{4,7123456}}} & b_{\text{5,2345716}} \\
 b_{\text{2,7123456}} & b_{\text{3,4567123}} & b_{\text{3,7123456}} & {b_{\text{4,2371456}}} & {b_{\text{4,7123456}}} & b_{\text{5,2345716}} \\
 b_{\text{3,4567123}} & b_{\text{3,2345671}} & b_{\text{4,2345671}} & {b_{\text{4,2371456}}} & {b_{\text{4,7123456}}} &{b_{\text{6,2345671}}}  \\
 b_{\text{3,2345671}} & b_{\text{3,7123456}} & b_{\text{4,2345671}} & {b_{\text{4,2371456}}} & {b_{\text{4,7123456}}} & {b_{\text{6,2345671}}}  \\
 b_{\text{3,4567123}} & b_{\text{3,7123456}} & b_{\text{4,2345671}} & {b_{\text{4,2371456}}} & {b_{\text{4,7123456}}} & {b_{\text{6,2345671}}} \\
\end{smallmatrix}
\right)\,.
\ee
The missing binomials associated to each piece of the refinement are
\be
\left(
\begin{matrix}
p_{156} p_{237} - p_{123} p_{567},& p_{146} p_{237} - p_{123} p_{467},& p_{145} p_{237} - p_{123} p_{457}\\
p_{137} p_{245} - p_{127} p_{345},& p_{156} p_{237} - p_{123} p_{567},& p_{146} p_{237} - p_{123} p_{467}\\
p_{137} p_{245} - p_{127} p_{345},& p_{167} p_{235} - p_{157} p_{236},& p_{167} p_{234} - p_{147} p_{236}\\
p_{137} p_{256} - p_{127} p_{356},& p_{137} p_{246} - p_{127} p_{346},& p_{137} p_{245} - p_{127} p_{345}\\
p_{145} p_{237} - p_{123} p_{457},& p_{167} p_{235} - p_{157} p_{236},& p_{167} p_{234} - p_{147} p_{236}\\
p_{167} p_{235} - p_{157} p_{236},& p_{167} p_{234} - p_{147} p_{236},& p_{157} p_{234} - p_{147} p_{235}\\
p_{137} p_{245} - p_{127} p_{345},& p_{167} p_{235} - p_{157} p_{236},& p_{167} p_{234} - p_{147} p_{236}
\end{matrix}
\right)
\ee
The second class has a representative with rays 
\be
\left\{b_{\text{5,7123564}},b_{\text{1,1234567}},b_{\text{4,7123456}},b_{\text{5,5671342}},b_{\text{4,5671234}},b_{\text{2,5671234}},b_{\text{4,5634712}},b_{\text{3,5671234}}\right\}\,,
\ee
which has the refinement
\be
\left(
\begin{array}{cccccc}
 b_{\text{1,1234567}} & b_{\text{2,5671234}} & b_{\text{4,7123456}} & b_{\text{4,5671234}} & b_{\text{5,7123564}} & b_{\text{5,5671342}} \\
 b_{\text{1,1234567}} & b_{\text{4,7123456}} & b_{\text{4,5634712}} & b_{\text{4,5671234}} & b_{\text{5,7123564}} & b_{\text{5,5671342}} \\
 b_{\text{1,1234567}} & b_{\text{3,5671234}} & b_{\text{4,7123456}} & b_{\text{4,5634712}} & b_{\text{4,5671234}} & b_{\text{5,7123564}} \\
 b_{\text{1,1234567}} & b_{\text{2,5671234}} & b_{\text{4,5634712}} & b_{\text{4,5671234}} & b_{\text{5,7123564}} & b_{\text{5,5671342}} \\
 b_{\text{1,1234567}} & b_{\text{2,5671234}} & b_{\text{3,5671234}} & b_{\text{4,5634712}} & b_{\text{4,5671234}} & b_{\text{5,7123564}} \\
\end{array}
\right)
\ee
and the missing binomials 
\be
\left(
\begin{matrix}
 p_{167}p_{345} - p_{157}p_{346}, & p_{167}p_{245} - p_{157}p_{246}\\
 p_{167}p_{345} - p_{157}p_{346}, & p_{147}p_{256} - p_{127}p_{456}\\
 p_{147}p_{356} - p_{137}p_{456}, & p_{147}p_{256} - p_{127}p_{456}\\
 p_{167}p_{345} - p_{157}p_{346}, & p_{156}p_{247} - p_{124}p_{567}\\
 p_{156}p_{347} - p_{134}p_{567}, & p_{156}p_{247} - p_{124}p_{567}
\end{matrix}
\right)\,.
\ee

Finally, we arrive at the $9$-vertex cones of which there are $7$ in a single dihedral class. A representative has the rays 
\be
\left\{b_{\text{4,2345671}},b_{\text{3,2345671}},b_{\text{5,2345716}},b_{\text{4,2371456}},b_{\text{4,4567123}},b_{\text{3,4567123}},b_{\text{5,4567231}},b_{\text{4,4523671}},b_{\text{2,2345671}}\right\}
\ee
and a refinement given by 
\be
\left(
\begin{array}{cccccc}
 b_{\text{3,4567123}} & b_{\text{3,2345671}} & b_{\text{4,2345671}} & b_{\text{4,2371456}} & b_{\text{4,4567123}} & b_{\text{4,4523671}} \\
 b_{\text{3,2345671}} & b_{\text{4,2345671}} & b_{\text{4,2371456}} & b_{\text{4,4567123}} & b_{\text{4,4523671}} & b_{\text{5,4567231}} \\
 b_{\text{2,2345671}} & b_{\text{3,2345671}} & b_{\text{4,2345671}} & b_{\text{4,2371456}} & b_{\text{4,4523671}} & b_{\text{5,4567231}} \\
 b_{\text{3,4567123}} & b_{\text{4,2345671}} & b_{\text{4,2371456}} & b_{\text{4,4567123}} & b_{\text{4,4523671}} & b_{\text{5,2345716}} \\
 b_{\text{4,2345671}} & b_{\text{4,2371456}} & b_{\text{4,4567123}} & b_{\text{4,4523671}} & b_{\text{5,2345716}} & b_{\text{5,4567231}} \\
 b_{\text{2,2345671}} & b_{\text{4,2345671}} & b_{\text{4,2371456}} & b_{\text{4,4523671}} & b_{\text{5,2345716}} & b_{\text{5,4567231}} \\
 b_{\text{2,2345671}} & b_{\text{3,4567123}} & b_{\text{4,2345671}} & b_{\text{4,4567123}} & b_{\text{4,4523671}} & b_{\text{5,2345716}} \\
 b_{\text{2,2345671}} & b_{\text{4,2345671}} & b_{\text{4,4567123}} & b_{\text{4,4523671}} & b_{\text{5,2345716}} & b_{\text{5,4567231}} \\
\end{array}
\right)\,,
\ee
with the missing binomials
\be
\left(
\begin{matrix}
p_{237}p_{456} - p_{236} p_{457}, & p_{145} p_{237} - p_{123}p_{457}, & p_{145} p_{236} - p_{123} p_{456}\\
p_{235}p_{467} - p_{234}p_{567},& p_{145}p_{237} - p_{123}p_{457},& p_{145}p_{236} - p_{123}p_{456}\\
p_{235}p_{467} - p_{234}p_{567},& p_{157}p_{234} - p_{147}p_{235},& p_{156}p_{234} - p_{146}p_{235}\\
p_{237}p_{456} - p_{236}p_{457},& p_{137}p_{245} - p_{127}p_{345},& p_{145}p_{236} - p_{123}p_{456}\\
p_{235}p_{467} - p_{234}p_{567},& p_{137}p_{245} - p_{127}p_{345},& p_{145}p_{236} - p_{123}p_{456}\\
p_{235}p_{467} - p_{234}p_{567},& p_{137}p_{245} - p_{127}p_{345},& p_{156}p_{234} - p_{146}p_{235}\\
p_{267}p_{345} - p_{245}p_{367},& p_{137}p_{245} - p_{127}p_{345},& p_{136}p_{245} - p_{126}p_{345}\\
p_{235}p_{467} - p_{234}p_{567},& p_{137}p_{245} - p_{127}p_{345},& p_{136}p_{245} - p_{126}p_{345}
\end{matrix}
\right)\,.
\ee

\pagebreak

\bibliographystyle{JHEP}
\bibliography{biblio}

\end{document}